\documentclass[aps,pra,12pt,onecolumn,showpacs,preprintnumbers,amsmath,amssymb,
floatfix,superscriptaddress]{revtex4-1}

\usepackage[utf8]{inputenc}
\usepackage{graphics}
\usepackage{graphicx}
\usepackage{amsmath}
\usepackage{setspace}
\usepackage{braket}
\usepackage{epstopdf}
\usepackage{comment}
\usepackage{hyperref}
\usepackage{bm}
\usepackage{xfrac}
\usepackage{textgreek}
\usepackage{xcolor}

\hypersetup{%
   pdfpagemode=None, 
   pdfstartpage=1,
   pdfmenubar=true,
   pdftoolbar=true,
   colorlinks = true,
   linkcolor=blue,
   citecolor=blue,
   urlcolor=blue,
   bookmarksopen=false
 }

\newcommand{\affA}{Van der Waals-Zeeman Institute, Institute of Physics,
University of Amsterdam, 1098 XH Amsterdam, Netherlands}
\newcommand{\affB}{QuSoft, Science Park 123, 1098 XG Amsterdam, the Netherlands}
\newcommand{\affC}{Physikalisch-Technische Bundesanstalt, Bundesallee 100, 38116 Braunschweig, Germany}
\newcommand{\affD}{Institut für Quantenoptik, Leibniz Universit\"{a}t Hannover, Welfengarten 1, 30167 Hannover, Germany}

\begin{document}

\title{Buffer gas cooling of ions in time-dependent traps using ultracold atoms}


\author{E.~Trimby}\affiliation{\affA}
\author{H.~Hirzler}\affiliation{\affA}
\author{H.~F\"urst}\affiliation{\affC}\affiliation{\affD}
\author{A. Safavi-Naini}\affiliation{\affA}\affiliation{\affB}
\author{R.~Gerritsma}\affiliation{\affA}\affiliation{\affB}
\author{R.~S.~Lous}\affiliation{\affA}

\begin{abstract}
For exploration of quantum effects with hybrid atom-ion systems, reaching ultracold temperatures is the major limiting factor. In this work, we present results on numerical simulations of trapped ion buffer gas cooling using an ultracold atomic gas in a large number of experimentally realistic scenarios. We explore the suppression of micromotion-induced heating effects through optimization of trap parameters for various radio-frequency (rf) traps and rf driving schemes including linear and octupole traps, digital Paul traps, rotating traps and hybrid optical/rf traps. We find that very similar ion energies can be reached in all of them even when considering experimental imperfections that cause so-called excess micromotion. Moreover we look into a quantum description of the system and show that quantum mechanics cannot save the ion from micromotion-induced heating in an atom-ion collision. The results suggest that buffer gas cooling can be used to reach close to the ion's groundstate of motion and is even competitive when compared to some sub-Doppler cooling techniques such as Sisyphus cooling. Thus, buffer gas cooling is a viable alternative for ions that are not amenable to laser cooling, a result that may be of interest for studies into quantum chemistry and precision spectroscopy.

\end{abstract}

\maketitle

\section{Introduction}
By immersing a trapped ion in a cold atomic bath, one creates a system that benefits both from the spatial localization and addressability of the ion, as well as from the scalability and ultracold temperatures of the atom cloud~\cite{Tomzareview,Harter2014CAIrev,Cote2016UHAI}. The result is a unique combination of two well-controllable platforms linked together by the intermediate range ion-atom interaction, with applications in impurity physics~\cite{Ratschbacher2013DSI,Schmid2010DCT,Furst2018DSI,Kollath2007STM,Astrak2021IPBEC,Dieterle2021TSI,Hirzler2020CNC,Zipkes2010TSI, Oghittu2021doa}, cold chemistry~\cite{Ratschbacher2012CCR,Haze2015CEC,Sikorsky2018SCA,Mohammadi2021LDC} and quantum technology~\cite{Bissbort2013ESS,Doerk2010AIQ}. Buffer gas cooling of trapped ions has long been used to obtain cold samples~\cite{Sauder1968TEC,Chen2014NGS,Zipkes2011KST,Meir2016DGS}. However, in efforts to reach quantized ion motion it has been outperformed by sub-Doppler laser cooling techniques, which have allowed small ion crystals to be cooled to their ground state of motion~\cite{Ioncooling,Morigi2000GSLC}. Meanwhile, developments in laser and evaporative cooling techniques have opened up the possibility to prepare long-lived, ultracold atomic gases below about 1\,$\mu$K, making them excellent buffer gases for cooling single ions. This progress has allowed buffer gas cooling to become a promising route towards ultracold atom-ion mixtures and exploring the realm of quantum effects within them.

Reaching the lowest possible temperatures in a hybrid atom-ion system is crucial to taking full advantage of the research opportunities it has to offer. In particular, an ion cooled to the regime of quantized motion would make an excellent probe to monitor interactions within the atomic bath and study impurity physics. Furthermore, cooling of both the atomic bath and the ion to atom-ion collision energies below the quantum limit would allow tuning of their interaction via $s$-wave Feshbach resonances~\cite{Idziaszek2011MCQ,Tomza2015CIY}, similar to those widely used in neutral atom systems~\cite{Chin2010FR} and recently observed in atom-ion systems~\cite{SchaetzFRs}. In previous work, we demonstrated that buffer gas cooling can be used to cool an atom-ion system to the edge of the $s$-wave collision energy, where the ion reached an average of 3.7 motional quanta~\cite{Nature}. However, efforts to reach deeper into the quantum regime or to buffer gas cool an ion to its motional ground state have remained elusive.

A major obstacle limiting buffer gas cooling of an ion in a radiofrequency (rf) trap is the heating effect of the time-dependent potential. In such a trap, energy stored in the ion's micromotion can be released during an atom-ion collision, heating the system~\cite{Cetina}. This competes with the desired cooling effects of the buffer gas collisions. As a result, the temperature of the ion after thermalization is always higher than that of the buffer gas in which it is immersed. Reduction of atom-ion collision energies to the quantum regime are therefore predominantly limited by the kinetic energy of the ion, when trapping in a rf ion trap. For this reason, minimizing the ion's excess micromotion (EMM) through compensation of experimental imperfections is critical to buffer gas cooling experiments~\cite{HAF}. Furthermore, choosing an atom-ion species combination with a large mass ratio greatly suppresses the effects of micromotion-induced heating~\cite{Cetina,Haze2018CDS}. Therefore, the coldest atom-ion collision energies and subsequent observation of quantum effects have been reported for Li/Yb$^+$~\cite{Nature} and Li/Ba$^+$~\cite{SchaetzFRs} systems. Attempts to avoid these micromotion heating effects altogether by trapping ions in other types of trap, for example optical traps, are ongoing, yet offer different challenges~\cite{arxivKarpa}. For an ion in a Paul trap, the question remains: what are the lowest temperatures achievable in a rf-trapped ion-atom system?

In this paper, we numerically simulate buffer gas cooling in time-dependent traps, showing that the optimization of trap parameters is crucial to reaching the lowest possible energies. We focus on the Li/Yb$^+$ species combination, and find that parameter optimization can allow cooling to collision energies approximately two times smaller than what has previously been achieved experimentally~\cite{Nature}. For these settings, we expect an occupancy of about one motional quanta on average, putting buffer gas cooling on par with sub-Doppler cooling techniques such as Sisyphus cooling~\cite{Ejtemaee20173DS}. We study the collision energies that can be achieved by buffer gas cooling in alternative rf traps including hybrid rf-optical traps, digital Paul traps, rotating rf-traps and hexapole and octupole traps. We find that all investigated traps can be used to achieve the same low collision energies as in a linear Paul trap. The only investigated traps that offer a clear advantage over the linear Paul trap are the rf-optical hybrid trap and multipole traps, in which even lower temperatures can be reached. However these depend on the use of ambitious experimental parameters and, in the case of the hybrid trap, suppression of photon scattering effects which could otherwise lead to heating of the atom cloud. All simulations show that within these time-dependent traps, we can reach the energy regime where a quantum mechanical treatment of the atom-ion interaction becomes necessary. Therefore, we also discuss how to describe the atom-ion collision within a quantum description and show that  the micromotion-induced heating survives.

The paper is organised as follows. In section~\ref{sec:method} we outline the method with which we classically simulate the buffer gas cooling process, indicating how we extract and compare the ion kinetic energies and atom-ion collision energies reached in these simulations. In section~\ref{sec:lin} we detail the time-dependent potential used to confine an ion in a linear Paul trap, and show simulation results for the ion energies that can be reached in such a trap with variable parameters. In section~\ref{sec:other}, we consider other types of time-dependent ion traps including digital, rotating and multipole traps, comparing buffer gas cooling simulations in each of these to the results of the linear Paul trap. We make a similar comparison in section~\ref{sec:hybridtrap}, now focusing on a hybrid optical-rf trap. Finally, in section~\ref{sec:quantum}, we present a quantum description of the atom-ion interaction to support our results also in the regime of low kinetic energies of the ion.

\section{Method}\label{sec:method}
For our investigation into buffer gas cooling, we simulate repeated collisions between a trapped ion and atoms from an ultracold cloud. During these simulations, there are two effects that dictate the motion of the colliding particles as they are propagated towards each other in time. These are the atom-ion interaction $V_\text{ai}(r)$, and the time-dependent potential used to trap the ion $V_\text{trap}(t,r)$. The latter is adjusted to describe the rf trap we wish to investigate. Our simulations follow the method outlined in ref.~\cite{HAF}, which has been shown to be in good agreement with experimental results~\cite{Nature}. We adapt this process to allow for a broader choice of $V_\text{trap}(t,r)$, corresponding to a range of rf ion traps. We study the role played by $V_\text{trap}(t,r)$, and its defining parameters, in dictating the properties of the simulated ion after buffer gas cooling. 

The atom-ion interaction consists of a long-range induced-dipole interaction, in which the charge of the ion polarizes the atom, giving rise to an attraction between them, as well as a short-range repulsion. The interaction is described in our simulation as
\begin{equation}\label{eq:atomionint}
  V_\text{ai}(r)=-\frac{C_\text{4}}{r^4}+\frac{C_\text{6}}{r^6}\,,
\end{equation}
with $r$ the separation of the colliding particles. Here, $C_4$ and $C_6$ are the long-range attraction and short-range repulsion interaction coefficients, respectively. The choice of $C_6$ dictates the point at which the atom-ion interaction becomes repulsive. It was shown in~\cite{HAF} that its exact value is unimportant as long as the $r^{-6}$ term starts to dominate at a short enough length scale of $\sim$1\,nm. Meanwhile, the value of $C_4$ depends on the polarizability of the species of buffer gas atoms. To describe buffer gas cooling of an ion to the mK range, the short-range repulsion is enough~\cite{Chen2014NGS}. However for cooling beyond this, the induced-dipole interaction has been shown to play a significant role~\cite{Cetina}, as it is the mechanism by which the ion can be pulled from the trap centre into a region where micromotion heating limits its final temperature. Both terms are therefore crucial to our simulation of atom-ion collisions at the $\mu$K energy level.

During each simulated atom-ion collision, we use a fourth-order step-size-adaptive Runge-Kutta algorithm to propagate one atom towards the confined ion, under the influence of $V_\text{trap}(t,r)$ and $V_\text{ai}(r)$. We assume the atom cloud is dilute enough to ignore non-binary collisions between the ion and the atoms. The ion's initial kinetic energy is set to 0\,K. The atom position is chosen at random on a sphere centred around the ion at the start of each collision. The sphere size is set to be large enough to accommodate large ion motion if and when the ion is heated.  The atom's velocity in the radial (angular) direction is also selected at random from a Weibull (Gaussian) distribution corresponding to a chosen buffer gas temperature in the $\mu$K range. The two particles are then propagated until the atom leaves the sphere. After each simulated collision we record the ion's average kinetic energy ($E_\text{kin}$) and motional properties. These are not reset after each collision. We investigate the final values achieved after the ion has thermalized, usually after some hundreds of collisions. To stay within a reasonable simulation time, the temperature of the buffer gas is set to 2\,$\mathrm{\mu K}$, which corresponds to the experimental values of ref.~\cite{Nature}. 

\begin{figure}[!t]
\centering
\includegraphics[width=20pc]{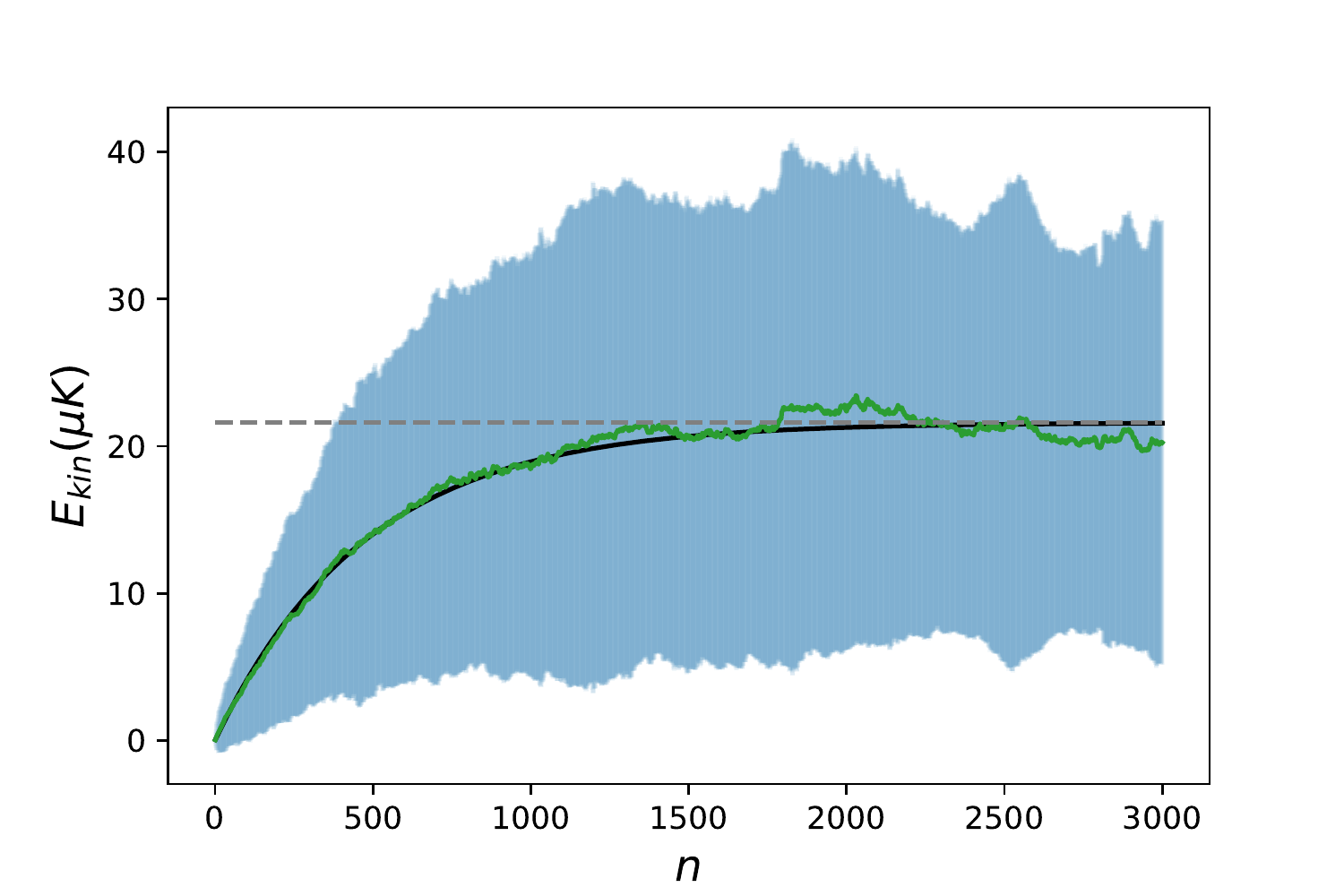}
\caption{Typical thermalization curve when simulating consecutive atom-ion collisions. The green line shows the average temperature reached after each collision, averaged over 221 runs of the thermalization process. The blue shaded area indicates the standard deviation about this average value and represents the width of the distribution of temperatures after the $n$th collision. The black line shows the fit of equation~\eqref{thermcurvefit} to the averaged data, with the grey dashed line indicating the $E_\text{kin,f}$ found from this fit.}
\label{Errorbars and fit}
\end{figure}

An example of the typical ion thermalization data obtained in our simulation is shown in Figure~\ref{Errorbars and fit}. For each simulation, we run at least 200 repeats of the thermalization process, averaging over them to find the mean ion energy reached after the $n$th collision (green line). The blue error bars correspond to the standard deviation in these energies. In general, higher ion energies are accompanied by larger error bars. This is because, dependent on the mass ratio of the atom-ion species combination, the ion energy statistics after multiple collisions follow either a  Tsallis or Maxwell-Boltzmann distribution~\cite{DeVoe2009PLD,Rouse2017SSE,Nature}. In both cases, the higher the mean ion energy, the larger the statistical spread of ion energies about this mean, as reflected in the error bars.
To extract the final steady-state ion energy, we perform a weighted least-squares fit to our data. The weighting of the fit is determined by the uncertainty at each data point. We fit according to the equation
\begin{equation}
\label{thermcurvefit}
    E_\text{kin,i}(n) = (E_\text{kin,f}-E_\text{kin,0})(1-e^{-n\Gamma})+E_\text{kin,0}\,.
\end{equation}
Here, $E_\text{kin,0}$ is the initial kinetic energy of the ion, $\Gamma$ indicates the rate of thermalization and $E_\text{kin,f}$ is the kinetic energy reached by the ion at steady state. The latter two are fitting parameters. In figure~\ref{Errorbars and fit}, the fit is shown by the  black curve, and the steady-state kinetic energy is indicated by the grey, dashed line. Note that $n$ in this plot includes both Langevin and glancing collisions, meaning that the value of $\Gamma$ extracted from the fit depends strongly on the initial atom-ion separation. However, within the range used in our simulation, this separation does not influence $E_\text{kin,f}$ as shown in ref.~\cite{HAF}.

We focus our simulations on the Li/Yb$^+$ species combination, due to its proved effectiveness in reaching $s$-wave collision energies in buffer gas cooling experiments~\cite{Nature}. The species choice has implications for the atom-ion interaction strength. Choosing $^6$Li/$^{171}$Yb$^+$ corresponds to setting $C_4=1.121\times 10^{-56}\,\mathrm{Jm}^4$. For the short-range repulsion we use $C_6=2.804\times 10^{-75}\,\mathrm{Jm}^6$. The induced-dipole interaction has a characteristic range $R^*=\sqrt{\frac{\mu C_4}{\hbar^2}}$, with a corresponding energy scale given by $E^*=\frac{\hbar^2}{2\mu(R^*)^2}$, which also represents the s-wave energy limit for the system. In a $^6$Li/$^{171}$Yb$^+$ system, these take the values 69.75\,nm and 8.58\,$\mu$K, respectively. To reach deep into the quantum regime with buffer gas cooling, the kinetic energy of the ion must be reduced such that the collision energy of the atom-ion system $E_\text{col}=\frac{\mu}{m_\text{i}}E_\text{kin,i}+\frac{\mu}{m_\text{a}}E_\text{kin,a}$ is significantly lower than $E^*$. Here, $\mu$ is the reduced mass. 

\section{The linear Paul trap}\label{sec:lin}
In an ideal linear Paul trap, the time-dependent potential can be written as
\begin{equation}\label{eq:vtrap_lin}
    V_\text{trap}(t,r) = \frac{-v_\text{dc}}{R_\text{dc}^2}\sum_{j=x,y,z}\alpha_jr_j^2 + \frac{v_\text{rf}}{2R_\text{rf}^2}\cos{(\Omega_\text{rf}t)}\sum_{j=x,y,z}\alpha_j'r_j^2\,,
\end{equation}
with $\alpha_x=\alpha_y\approx-\alpha_z/2=-0.5$ and $\alpha_x'=-\alpha_y'=1$, $\alpha_z'=0$~\cite{Leibfried,HAF}. Here, $v_\text{dc}$ and $v_\text{rf}$ correspond to the direct-current (dc) and radio-frequency (rf) voltages applied to the trap electrodes. Meanwhile, $R_\text{dc}$  and $R_\text{rf}$ describe the fixed geometry of the trap and are approximately half the separation between opposing dc and rf electrodes, respectively. The frequency at which the rf voltage oscillates is given by $\Omega_\text{rf}$. The two terms of equation~\eqref{eq:vtrap_lin} are commonly referred to as $V_\text{trap}^\text{dc}(r)$ and $V_\text{trap}^\text{rf}(t,r)$, respectively. The resulting ion motion is described by the Mathieu equations, $\ddot{r}_j=-(a_j+2q_j\cos{(\Omega_\text{rf}t)})\frac{\Omega_\text{rf}^2}{4}r_j$,
with $a_j$, $q_j$ the stability parameters of the ion trap for all three directions ($j=x,y,z$). The stability parameters can be defined in terms of the trap geometry and voltage signals as $a_{x,y}\approx-a_z/2=4ev_\text{dc}/(m_\text{i}R_\text{dc}^2\Omega_\text{rf}^2)$, and $q\equiv q_x=-q_y=2ev_\text{rf}/(m_\text{i}R_\text{rf}^2\Omega_\text{rf}^2)$, $q_z$=0, with $e$ the charge of the ion. The stability parameters combined with $\Omega_\text{rf}$ determine the frequency $\omega_{\text{sec},j}\approx\frac{\Omega_\text{rf}}{2}\sqrt{a_j+\frac{q_j^2}{2}}$ of the trapped ion's secular motion. 

In an ideal trap, the ion's energy is mainly given by the kinetic energy, which is averaged over one period of the ion's secular motion. It is defined as
\begin{equation}\label{eq:ionsteadystateKE}
    \bar{E}_\text{kin,j}\approx\frac{1}{4}m_\text{i}A_j^2\left(\omega_{\text{sec,}j}^2+\frac{1}{8}q_j^2\Omega_\text{rf}\right)\,,
\end{equation}
and depends on the secular frequency and amplitude $A_j$ of the ion motion, and the trap parameters $q$ and $\Omega_\text{rf}$. The latter two can be easily changed in experiments by changing the frequency and amplitude of the rf signal applied to the trap electrodes, and will be studied in our work. Typical numbers used in our Li/Yb$^+$ experiment are $q=0.5$, $\Omega_\text{rf}=2\pi\times 2$\,MHz and $v_\text{dc}$ up to 6.3\,V, corresponding to $\omega_{\text{sec,}x}$ and $\omega_{\text{sec,}y}$ up to $2\pi\times$350\,kHz, and $\omega_{\text{sec,}z}$ up to $2\pi\times$120\,kHz. The size of our trap is $R_\text{dc}=5$\,mm and $R_\text{rf}=1.5$\,mm, as outlined in ref.~\cite{setup}. 

The ion's kinetic energy is connected to the number of excited motional quanta $n_j$ of the ion in each direction $j$ by the following equations
\begin{equation}
    T_{\text{sec,}j} = \hbar \omega_{\text{sec,}j}\left(n_j+\frac{1}{2}\right)/k_\text{B}\,,
\end{equation}
and
\begin{equation}
    \sum_{j=x,y,z}\bar{E}_{\text{kin},j}\approx\frac{k_\text{B}}{2}\left(\sum_{j=x,y,z} T_{\text{sec,}j}+T_{\text{sec,}x}+T_{\text{sec,}y}\right)\,.
\end{equation}
Here $T_{\text{sec,}j}$, represents the secular temperature in direction $j$ and $k_\text{B}$ is the Boltzmann constant. 
The kinetic energy contains both secular motion (first term) and intrinsic micromotion (second and third terms) contributions~\cite{Berkeland}. Intrinsic micromotion (IMM) is driven by the rf field, and so depends on the radial temperature of the ion only. In our simulations, we focus on the average number of excited motional quanta in the radial direction ($\bar{n}=(n_x+n_y)/2$), as the axial motion of the ion is largely unaffected by variations in our parameters of interest, $q$ and $\Omega_\text{rf}$.

 In addition to $q$ and $\Omega_\text{rf}$, the ion's kinetic energy is also influenced by trap imperfections, which can lead to excess micromotion (EMM). Although trap-dependent, in a realistic simulation of the linear Paul trap these effects should be included~\cite{HAF,Nature}. There are three forms of EMM which all scale differently with $q$ and $\Omega_\text{rf}$. Axial EMM arises from the pickup of the rf signal on the dc electrodes, and contributes a kinetic energy ($E_\text{ax}$) that scales with $q^2$ and $\Omega_\text{rf}^{-2}$. Phase EMM, $E_\text{ph}$, results from a phase difference between the rf voltage signals on opposing electrodes, and scales with $q^2$ and $\Omega_\text{rf}^4$. Finally, $E_\text{rad}$ scales with $q^{-2}$ and $\Omega_\text{rf}^{-2}$, and arises when stray electric fields in the radial direction of the trap displace the ion from the trap centre. The ion's total kinetic energy is therefore
\begin{equation}
\label{eq:totalKEfit}
    E_\text{kin}=\sum_{j=x,y,z}\bar{E}_{\text{kin},j}+b E_\text{rad}+c E_\text{ax}+ d E_\text{ph} + g\,.
\end{equation}
Here, $b, c, d$ are fitting parameters indicating the contributions of the experiment-specific energies $E_\text{rad}$, $E_\text{ax}$ and $E_\text{ph}$. The fitting parameter $g$ is added when an overall offset of $E_\text{kin}$ needs to be taken into account. In experiments, we can use ion position tracking or Ramsey microwave spectroscopy to measure stray fields to an accuracy of $\sim$0.01\,Vm$^{-1}$~\cite{Nature}. It is feasible that radial stray fields can be compensated down to 0.05\,Vm$^{-1}$. Note that the finite size of the trap can lead to $q_z \neq 0$, which can give rise to additional EMM. However, since $q_z$ is much smaller than $q$ and its EMM effects can be minimised by selection of an appropriate ion position along the trap axis, we assume it is small enough to be ignored in our simulations.

We investigate the values that $E_\text{kin}$ and $\bar{n}$ can reach after buffer gas cooling by simulating collisions between $^6$Li atoms and an $^{171}$Yb$^+$ ion confined in a linear Paul trap with varying levels of EMM. The temperature of the Li cloud is set to 2\,$\mathrm{\mu K}$. We vary $q$ and $\Omega_\text{rf}$ independently, to explore their impact, setting $\Omega_\text{rf}=2\pi\times 2$\,MHz while $q$ is varied and $q=0.219$ while $\Omega_\text{rf}$ is varied. The simulated trap shares its geometry with that of ref.~\cite{setup} and has $v_\text{dc}=0.787$\,V such that $\omega_{\text{sec,}z}=2\pi\times 4.24$\,kHz.  We include experimentally feasible levels of axial and phase EMM corresponding to an axial rf pickup of 2.02\,Vm$^{-1}$ ($E_\text{ax}=42.4$\,$\mu$K) and a phase of 0.041\,millirad ($E_\text{ph}=9.18$\,$\mu$K) when $q=0.219$ and $\Omega_\text{rf}=2\pi\times 2$\,MHz~\cite{HAF,Harter,setup}.  
We also vary the strength of the radial stray field around experimentally realistic values. 

\begin{figure}[!t]
\centering
\includegraphics[width=37pc]{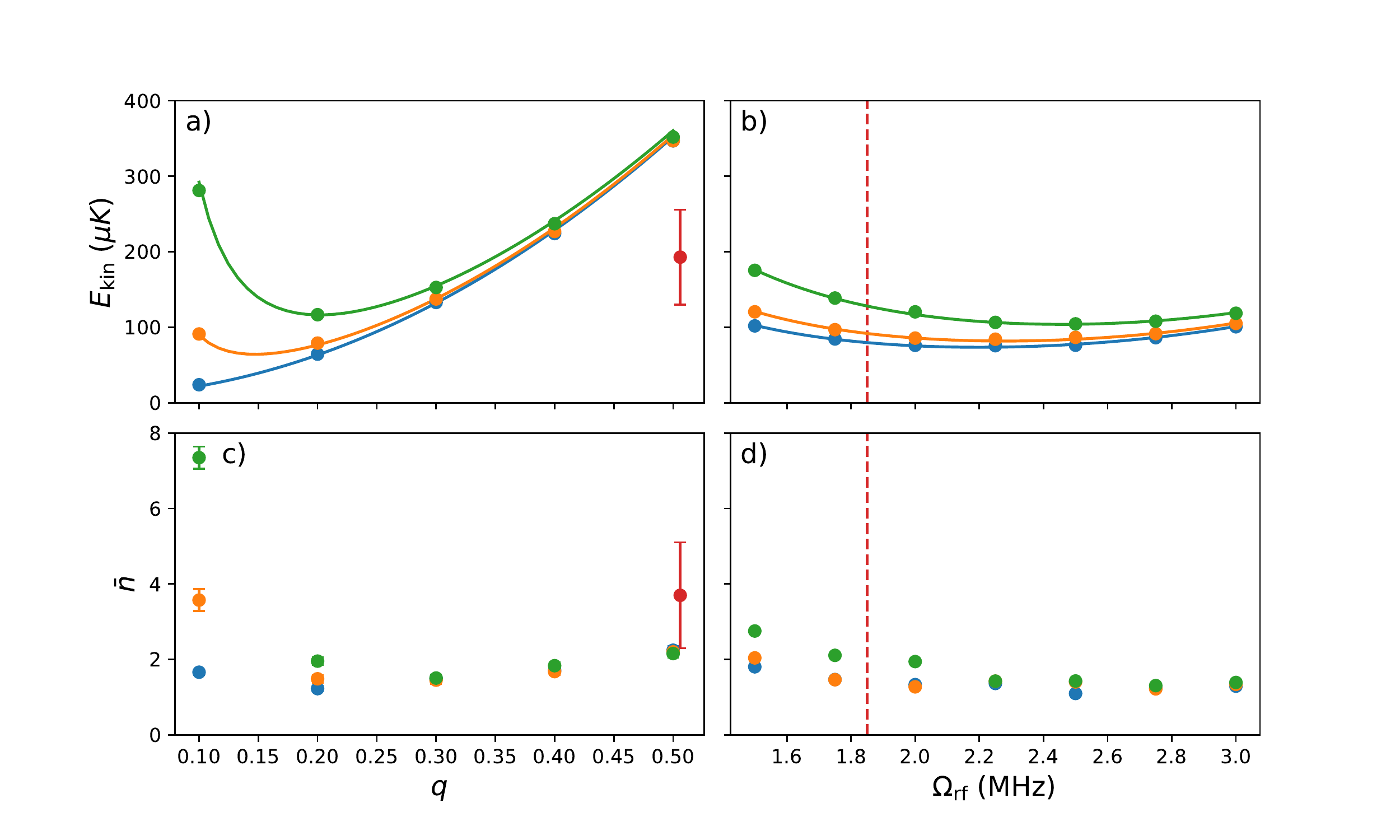}
\caption{Calculated ion energy and motion after buffer gas cooling in a linear Paul trap. The ion's steady-state kinetic energy $E_\text{kin}$ and number of excited motional quanta $\bar{n}$ are shown for cooling inside a trap of variable stability parameter $q$ and radio frequency $\Omega_\text{rf}$. Blue, orange and green data corresponds to a trap with a radial stray field of strength 0\,Vm$^{-1}$, 0.05\,Vm$^{-1}$ and 0.1\,Vm$^{-1}$, respectively. Red data points in \textbf{a} and \textbf{b} show experimental results from ref.~\cite{Nature}, while the red dashed lines in \textbf{c} and \textbf{d} indicate the value of $\Omega_\text{rf}$ used in the same experiment. The lines in \textbf{a} and \textbf{b} correspond to equation~\eqref{eq:totalKEfit}, fit to the data using fitting parameters \(b, c, d, g\) and \(A_{x,y,z}\). The fit values of these parameters can be found in Appendix A. Errors are standard errors and are sometimes smaller than the point size.}
\label{fig:qdrlinPaul}
\end{figure}

Our  simulations show that the optimal values of $q$ and $\Omega_\text{rf}$ for buffer gas cooling depend on the radial stray fields present in our system.  This is shown in figure~\ref{fig:qdrlinPaul}. Blue, orange and green data corresponds to a trap with a radial stray field of strength 0\,Vm$^{-1}$, 0.05\,Vm$^{-1}$ and 0.1\,Vm$^{-1}$, respectively. Figures~\ref{fig:qdrlinPaul}\textbf{a} and \textbf{b} show how the ion energy $E_\text{kin}$ varies with $q$ and $\Omega_\text{rf}$. In general, the optimal trap parameters for minimising the ion energy take non-zero values, and depend on the strength of the radial stray field. During atom-ion collisions, the trap's rf field does work to heat the ion mediated by the long-range attractive atom-ion interaction~\cite{Cetina}. This heating competes with buffer gas cooling and leads to a trade-off between two effects. On the one hand, high values of $q$ and $\Omega_\text{rf}$ provide tight ion confinement, meaning that the pull from the atoms does not significantly alter its motion. On the other hand, low values of $q$ and $\Omega_\text{rf}$ will decrease the energy of the rf electric field, lowering the energy that it can donate to the ion during an atom-ion collision. The optimal trap parameters therefore provide a trade-off between these effects. Moreover, EMM provides additional trap-parameter-dependent heating. The higher the field-strength of stray radial electric fields that additionally pull the ion from the Paul trap centre, the higher the favoured optimal values of $q$ and $\Omega_\text{rf}$. Appropriate selection of $q$ is particularly important for optimal buffer gas cooling. This can be seen by the steeper slope of the fit lines in Fig~\ref{fig:qdrlinPaul}\textbf{a}, compared to \textbf{b}.

The lowest achievable kinetic energy depends significantly on the possibility to reduce electric stray fields and optimal choice of $q$ and $\Omega_\text{rf}$. For $E_\text{kin}$ we find optimal values by fitting the data points using equation~\eqref{eq:totalKEfit}. The fit values can be found in Appendix A. For the chosen parameters, the lowest ion kinetic energy achievable by buffer gas cooling in our Paul trap is $64(1)\,\mu$K, which can be achieved using trap parameters $q=$0.147 and $\Omega_\text{rf}=2.28$\,MHz. In a 2\,$\mu$K atom cloud, this corresponds to $E_\text{col}=0.59\times E^*$, significantly below the $s$-wave limit. It is also two times lower than the lowest collision energy achieved in previous experiments~\cite{Nature}, which is indicated by the red data point in \textbf{a}. This experiment yielded collision energies $1.15\times E^*$ by buffer gas cooling an ion to kinetic energies of 193\,$\mu$K. The lowest achievable ion kinetic energies indicated by our simulation are therefore roughly 3 times colder than what has been achieved in experiments. Note that the discrepancy between the experimental data point and the simulation predictions can be explained by an additional cooling step of adiabatic decompression that was used in the plotted experiment, but is not included in our simulation. Our results suggest that with careful parameter optimization, this final step is not necessary for reaching low temperatures. The value 64\,$\mu$K is a lower limit on what can be achieved for the ion in this setup with an atomic bath of 2\,$\mu$K. Experiments will be limited to higher energies due to background heating from technical noise~\cite{setup}, which varies across experiments and is not included in our simulations. Quantum effects may also play a role in reducing the rate of buffer gas cooling at very low temperatures, as the rate of atom-ion collisions is expected to be lower once $E_\text{col}$ reaches below the $s$-wave energy~\cite{Nature}. More details on a quantum treatment of the atom-ion interaction can be found in section~\ref{sec:quantum}.

Depending on the stray fields, $\bar{n}$ is minimized for different values of the trap parameters to $E_\text{kin}$. This can be seen when comparing figure~\ref{fig:qdrlinPaul}\textbf{a} and \textbf{b} to \textbf{c} and \textbf{d}. Each of these variables is affected differently by sources of EMM. Therefore, to optimize buffer gas cooling experiments, one should consider whether to prioritize reaching quantized ion motion or $s$-wave collision energies to inform the choice of trap parameters. The lowest average number of excited motional quanta to which the ion can be cooled by an atom cloud at 2\,$\mu$K in our setup is 0.92(7), at a stability parameter $q=0.26$. This is 4 times smaller than what has been achieved in buffer gas cooling experiments, as indicated by the red data point in \textbf{c}. It allows us to reach quantized ion motion, making buffer gas cooling competitive with some sub-Doppler laser cooling techniques, such as Sisyphus cooling~\cite{Ejtemaee20173DS}. However, it is not capable of reaching $\bar{n}\ll1$, as is possible using, for example, resolved sideband cooling~\cite{Ioncooling}.

\section{Other types of rf ion traps}\label{sec:other}
Since the ion's final kinetic energy is limited by the trap parameters and excess micromotion in a linear Paul trap, this raises the question of whether other trap configurations are better suited for buffer gas cooling. Therefore, in this section, we simulate a digital ion trap (DIT), multipole ion trap (MIT) and rotating ion trap (RIT). We investigate how the dependence of $E_\text{kin}$ on the trap parameters may differ for different traps and, where relevant, consider the roles played by new parameters that are unique to certain traps.

\subsection{Digital ion trap}\label{subsec:dit}
The DIT is a modification of the previously-discussed linear Paul trap, whereby the rf voltage is driven \textit{digitally} (stepwise constant) instead of \textit{analogue} (sinusoidally). The switching of the voltage leads to times where the ion dynamics are that of a free particle as the rf voltage is zero and this might lead to more efficient buffer gas cooling in these periods. However, collisions occurring during a voltage switch will experience a violent change in potential, which may result in larger ion heating than in the traditional analogue Paul trap. 

Similar to equation~\eqref{eq:vtrap_lin}, the time-dependent potential of a DIT can be described by
\begin{equation}\label{eq:DITvoltage}
    V_\text{trap}(t,r)=V_\text{trap}^\text{dc} (r) + \frac{v_\text{rf}}{2R_\text{rf}^2}P_\text{DIT}(t)\sum_{j=x,y,z}\alpha_j'r_j^2\,.
\end{equation}
Here, the variables are equivalent to those in equation~\eqref{eq:vtrap_lin} apart from the shape of the rf signal $P_\text{DIT}(t)$, which is defined as
\begin{equation}\label{eq:DITpulse}
    P_\text{DIT}(t)=\begin{cases}
      1 & \text{when } t_{r}\leq \tau \, T/2\,,\\
      0 & \text{when } \tau \, T/2<t_{r}\leq(1-\tau)\, T/2\,,\\
      -1 & \text{when } (1-\tau)\, T/2<t_{r}\leq(1+\tau)\, T/2\,,\\
      0 & \text{when } (1+\tau)\, T/2<t_{r}\leq(2-\tau)\, T/2\,,\\
      1 & \text{when } (2-\tau)\, T/2<t_{r}\leq T\,.
    \end{cases}  
\end{equation}
Here, $T=2\pi/\Omega_\text{rf}$ is the period of the signal, $t_r$ is the remainder $t =\kappa\times T +t_r$~\cite{Deb}, with $\kappa$ an integer, and $\tau$ is the fractional pulse width. The digital voltage driving is shown in Figure~\ref{Tkin vs q,T DIT, pulse}\textbf{a}. For a DIT, $\tau$ adds  an additional trap parameter that affects the stability of the trap~\cite{KjaergaardDrewsen}. Approximating the digital signal with the first term of the corresponding Fourier series~\cite{Bandelow}, the $q$ parameter for a DIT becomes $q_{x,y,\text{DIT}}= 16ev_\text{rf}\sin{(\tau\pi)}/(\pi m_\text{i} R^2_\text{rf}\Omega_\text{rf}^2)$.

 We investigate how $E_\text{kin}$ varies when buffer gas cooling an ion in a DIT, dependent on the temperature of the atomic bath $T_\text{a}$, and the trap parameters $q$ and $\tau$. The results can be seen in figure~\ref{Tkin vs q,T DIT, pulse}\textbf{b} and \textbf{c}. In these simulations, we use the same geometry, trap parameters and atomic bath temperature as in the linear Paul trap simulations, except where otherwise stated. In figure~\ref{Tkin vs q,T DIT, pulse}\textbf{b}, we choose $\tau=0.285$ and $v_\text{rf}=34.467$\,V such that we have $q=0.219$, to match the previous simulations. In figure~\ref{Tkin vs q,T DIT, pulse}\textbf{c}, we vary $q$ and $\tau$ while keeping the trap geometry constant. For a fixed $\tau$, the variation in $q$ corresponds only to a change in the voltage of the digital pulses $v_\text{rf}$. The black lines show the trends as given by an analogue Paul trap with the same $q$, $\Omega_\text{rf}$ and EMM parameters. Radial stray field strengths are set to 0 or 0.1\,Vm$^{-1}$ in these simulations, while sources of axial and phase MM are not included.

Our simulation shows that buffer gas cooling using a digital rf driving results in equal ion temperatures to those in an analogue linear Paul trap. In figure~\ref{Tkin vs q,T DIT, pulse}\textbf{b}, DIT results are in excellent agreement with the analogue Paul trap trend line. Therefore, for the chosen $\tau$ and $q$ of this plot, the energies reached in the DIT are identical to those of an analogue trap, irrespective of the atomic bath temperature. In figure~\ref{Tkin vs q,T DIT, pulse}\textbf{c}, we see that in a DIT of variable $q$, the data points corresponding to the lowest $E_\text{kin}$ also lie on the trend line of the analogue trap. The range of $q$ for which the DIT and analogue trends match is affected by the fractional pulse width $\tau$. While $\tau$ provides an extra degree of freedom for tuning the trap stability, its tunability does not lead to lower ion kinetic energies than are achievable with an analogue Paul trap. The fact that the DIT results are so similar to those of the analogue Paul trap suggests that the time scale of an atom-ion collision is longer than the period of the rf trap, such that the shape of the rf voltage signal during the collision period is unimportant~\cite{Cetina}.

\begin{figure}[!t]
\centering
\includegraphics[width=37pc]{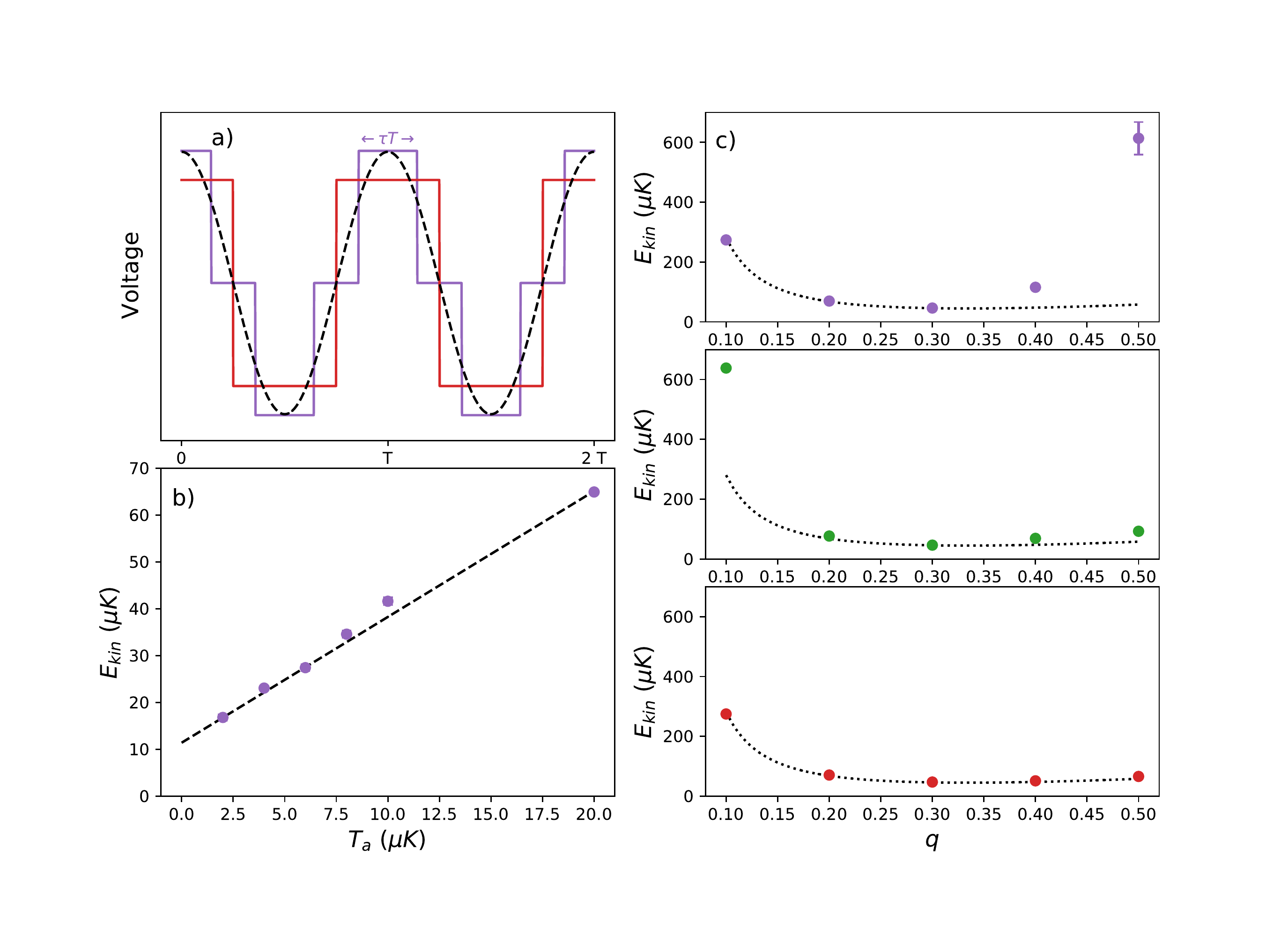}
\caption{Ion energy after buffer gas cooling in a DIT. \textbf{a} Analogue (black, dashed) versus digital (purple and red) driving of the rf voltage. The fractional pulse width $\tau$ defines the driving. Digital signals are shown for $\tau=0.285$ (purple) and $0.5$ (red). \textbf{b} The kinetic energy reached for variable buffer gas temperatures, for $q=0.219$ and $\tau=0.285$. \textbf{c} The kinetic energy reached for variable trap parameter $q$ for $\tau=0.285$ (purple), $0.4$ (green) and $0.5$ (red). The black dotted lines represent the trend in an analogue linear Paul trap. Errors are standard errors and are sometimes smaller than the point size.}
\label{Tkin vs q,T DIT, pulse}
\end{figure}

\subsection{Rotating ion trap}\label{subsec:rot}
In a RIT, ion confinement is provided by a rotating potential, instead of the oscillating potential typical for linear Paul traps.The ion motion is described by four compoments, instead of the infinite sum of Fourier components needed for a linear Paul trap. Since fast MM plays an important role in atom-ion collisions, we may therefore expect a different outcome for buffer gas cooling when comparing the two traps.

A RIT time-dependent trapping potential can be written as~\cite{HasegawaBollinger,rotsaddle}
\begin{equation}\label{eq:vtrap_rot}
    V_\text{trap}(t,r) = V_\text{trap}^\text{dc} (r) + \frac{v_\text{rf}}{2R_\text{rf}^2}[(r_x\cos{\Omega_\text{rf}t}-r_y\sin{\Omega_\text{rf}t})^2-(r_x\sin{\Omega_\text{rf}t}+r_y\cos{\Omega_\text{rf}t})^2]\,.
\end{equation}
Here, variables are equivalent to those in equation~\eqref{eq:vtrap_lin}, yet the rf driving part has changed. The resulting ion motion is then $\ddot{r}_x=-(a_xr_x+2q_x(r_x\cos{\Omega_\text{rf}t}-r_y\sin{\Omega_\text{rf}t}))\frac{\Omega_\text{rf}^2}{4}$, $\ddot{r}_y=-(a_yr_y+2q_y(r_y\cos{\Omega_\text{rf}t}+r_x\sin{\Omega_\text{rf}t}))\frac{\Omega_\text{rf}^2}{4}$, $\ddot{r}_z=-a_zr_z\frac{\Omega_\text{rf}^2}{4}$. An important consequence of this is that the ion's motion is now circular, instead of oscillating, and consists of four Fourier components: two secular motions and two IMM components. The frequency of the secular motion in the RIT can be written to first order in $q$ as
\begin{gather}
    \omega_{\text{sec,RIT}}\approx \pm\frac{\Omega_\text{rf}}{2}\sqrt{a+q^2}\,.
\end{gather}
Here $a=a_x=a_y$, and $q=q_x=-q_y=2ev_\text{rf}/(m_\text{i}R^2_\text{rf}\Omega_\text{rf}^2)$. Note that if $a$ is very small, the secular frequency differs from that of a linear ion trap by a factor $\sqrt{2}$. Therefore, the ion confinement in a RIT is twice as tight as in a linear trap with the same trap parameters. 

\begin{figure}[!t]
\centering
\includegraphics[width=37pc]{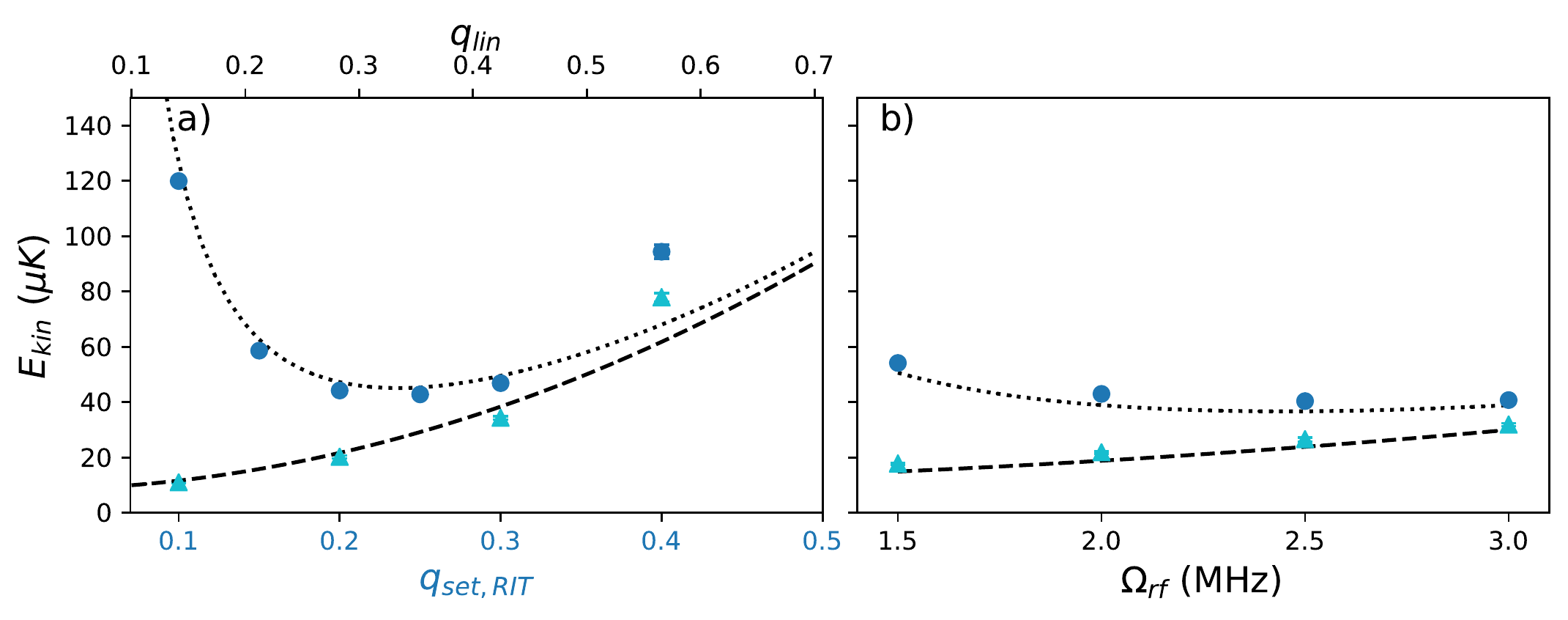}
\caption{Ion energy after buffer gas cooling in a RIT. Results are shown for traps of various $q$ (\textbf{a}) and $\Omega_\text{rf}$ (\textbf{b}), with (circles) and without (triangles) a radial stray field of 0.1V/m. The black dotted and dashed lines represent the trend in a linear Paul trap with similar trap parameters. Note that the black lines have been adjusted according to the relation $q_\text{eff,RIT}=\sqrt{2}q_\text{lin}$ (see text). Errors are standard errors and are sometimes smaller than the point size.}
\label{Tkin vs q dr ROT}
\end{figure}

We simulate buffer gas cooling of an ion in a RIT of variable $q$ and $\Omega_\text{rf}$, and explore how the dependence of $E_\text{kin}$ on these parameters compares to the case of the linear Paul trap. Results can be seen in figure~\ref{Tkin vs q dr ROT}. We set the same ion trap and atom cloud parameters as in our linear trap simulations. We simulate radial stray field strengths of 0 and 0.1\,Vm$^{-1}$ and do not include sources of axial and phase EMM. We set the stability parameter $q_\text{set,RIT}$ in terms of the trap geometry and driving signal as $q_\text{set,RIT}=q_\text{lin}=2ev_\text{rf}/(m_\text{i}R^2_\text{rf}\Omega_\text{rf}^2)$, which is the same definition as used in the linear Paul trap. However, due to the rotation of the potential, its effect on the ion energy differs from that of $q_\text{lin}$. We simulate buffer gas cooling in a RIT with and without a stray electric field of 0.1\,Vm$^{-1}$, as shown by the blue circles and triangles in figure~\ref{Tkin vs q dr ROT}. The data is compared to the equivalent linear trap trends, represented by the black dotted and dashed lines.

We find good agreement between the RIT data and the linear Paul trap trend at low $q$ when using $q_\text{eff,RIT}=\sqrt{2}q_\text{lin}$ to rescale the latter. Here, $q_\text{eff,RIT}$ refers to the effective stability parameter of the RIT and $q_\text{lin}$ is that of the linear trap. In figure~\ref{Tkin vs q dr ROT}\textbf{a} the x-axis of the black trend line is rescaled by $\sqrt{2}$ to obtain the match with the RIT data at low $q$, to within 3\,$\mu$K. The black line in \textbf{b} is also in good agreement with the RIT data when we use the same scaling of $q_\text{lin}$. The $E_\text{kin}$ dependence on $\Omega_\text{rf}$ and $q$ after this adjustment is thus similar for both traps, and significantly lower energies cannot be reached in the RIT. In figure~\ref{Tkin vs q dr ROT}\textbf{a}, the minimum $E_\text{kin}$ is achieved in the low-$q$ region. Only at high $q$ do the linear and RIT data differ significantly, and in this limit $E_\text{kin}$ is higher in the RIT. Note that the energies reached in this graph cannot be compared to those in figure~\ref{fig:qdrlinPaul}, as the latter includes sources of axial and phase MM, making the system inherently hotter. We conclude that although the $q_\text{set,RIT}$ we set in the simulation is defined identically in $v_\text{rf}$, $\Omega_\text{rf}$ and $R_\text{rf}$ to $q_\text{lin}$, the effect of the rotating potential is that of a linear trap that is a factor $\sqrt{2}$ higher in $q$. In a RIT, we can therefore use a lower-amplitude voltage signal $v_\text{rf}$ to achieve similar ion temperatures as in a linear Paul trap.

\subsection{Multipole ion trap}
The benefits of using a MIT for buffer gas cooling of molecular and atomic ions on the $\sim$\,K-mK level has been extensively explored~\cite{Wester2009RMT,Asvany2009NSK,Notzold2020TMI}. It has been shown that an increased number of trap poles minimises rf heating effects, allowing more efficient buffer gas cooling than would otherwise be achievable, as well as stable cooling of systems with a smaller atom-ion mass ratio than is possible in quadrupole traps~\cite{HoltkemeierPRL}. Here we investigate whether the same benefits extend to cooling of the high mass-ratio Li/Yb$^+$ system, beyond the lowest possible energies that can be reached in a quadrupole trap.

In a MIT, the four electrodes of the Paul trap are replaced by a higher number of electrodes, which are used to create an rf potential with $N$ poles. This requires $N$ ground and $N$ rf electrodes. For a Paul trap $N=2$, while a hexapole trap uses $N=3$ and an octupole trap has $N=4$. The resulting potential can be written as~\cite{MultiTextbook}
\begin{equation}
    V_\text{trap}(t,r)=V_\text{trap}^\text{dc} (r) +  \frac{v_\text{rf}}{2R_\text{rf}^2}r_\perp^N\cos{N\phi}\,.
\end{equation}
Here we use cylindrical coordinates $r_\perp=\sqrt{r_x^2+r_y^2}$ and $\phi=\tan^{-1}(r_y/r_x)$. The resulting motion in each direction $j=x,y,z$ will depend on $N$ and can be expressed as $\ddot{r_x}=-(a_x+\frac{N}{(N-1)}\eta_x\cos{\Omega_\text{rf}t})\frac{\Omega_\text{rf}^2}{4}r_\perp\cos{[(N-1)\phi]}$, $\ddot{r_y}=-(a_y+\frac{N}{(N-1)}\eta_y\cos{\Omega_\text{rf}t})\frac{\Omega_\text{rf}^2}{4}r_\perp\sin{[(N-1)\phi]}$ and  $\ddot{r}_z=-a_zr_z\frac{\Omega_\text{rf}^2}{4}$. Here, all parameters are defined as in equation~\eqref{eq:vtrap_lin}, except now the rf-dependent motion is written in terms of a generalized stability parameter $\eta_j$. We define this in terms of $q$ as
\begin{equation}
    \eta_j(r_\perp)=q_j(N-1)\left(\frac{r_\perp}{R_\text{rf}}\right)^{(N-2)}\,,
\end{equation}
such that $\eta_x=-\eta_y=q$ when $N=2$. Here, $q=q_x=-q_y=2ev_\text{rf}/(m_\text{i} R^2_\text{rf}\Omega_\text{rf}^2)$ and $q_z=0$. For $N>2$, $\eta_j$ has a $r_\perp$-dependence that is not present in a quadrupole trap. Depending on its energy, the ion oscillates around the trap axis with an amplitude $A_\perp$. Therefore, in a MIT used for buffer gas cooling, the stability parameter $\eta_j$ changes throughout the cooling process as the ion's energy decreases. In experiments, to ensure that $\eta(A_\perp)$ is kept roughly constant, changes to $A_\perp$ can be compensated by ramping the other trap parameters, for instance $v_\text{rf}$ can be easily ramped.

We simulate buffer gas cooling of an ion in a hexapole ($N=3$) and octupole ($N=4$) trap of variable stability. We use the same trap parameters as in the Paul trap from section~\ref{sec:lin}, except for $v_\text{rf}$, which we use to tune the trap stability $\eta_j$. We set $v_\text{rf}$ to define $\eta_j(r_\perp)=\eta$ at a chosen ion position $r_\perp=100$\,nm. We repeat the simulations with and without a radial stray electric field of 0.05\,Vm$^{-1}$. Attempts were made to simulate the process at stronger fields, however for $N>2$ this frequently resulted in a very unstable trap, as the ion was pulled to high $r_\perp$. Propagating the ion under these conditions is beyond the capability of our simulation. Similarly, for simulations with $N=4$ and $\eta>0.2$, frequent stability problems were encountered, making it difficult to repeat the simulations. Note that these simulations do not include sources of axial and phase EMM.

\begin{figure}[!t]
\centering
\includegraphics[width=37pc]{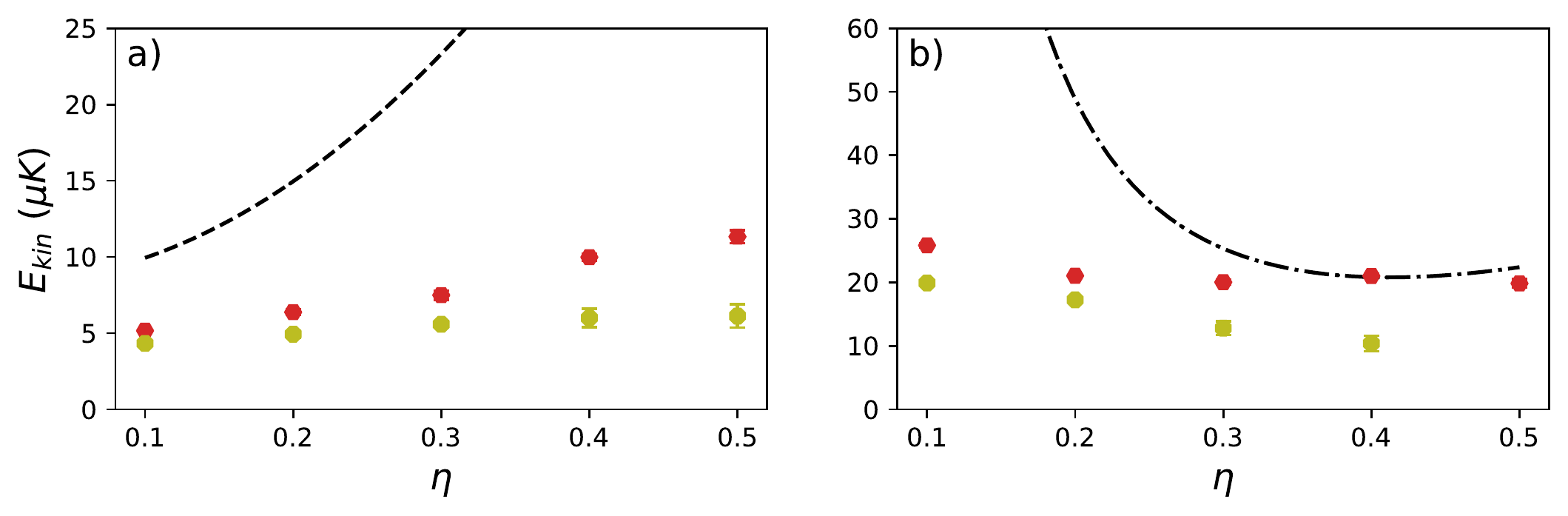}
\caption{Ion energy after buffer gas cooling in a multipole ion trap. Results are shown for an ion in a hexapole (red) and octupole (olive) trap, without (\textbf{a}) and with (\textbf{b}) a radial stray field of 0.05\,V/m. The dotted and dash-dotted lines are the fit to the data from the corresponding quadrupole trap without and with the radial stray field, respectively. Errors are standard errors and are sometimes smaller than the point size.}
\label{Tkin vs q dr MULTI}
\end{figure}

The greater the number of poles of the MIT, the more efficient the buffer gas cooling process. Figure~\ref{Tkin vs q dr MULTI} shows the variation of $E_\text{kin}$ with $\eta$ for each trap, with (\textbf{b}) and without (\textbf{a}) stray electric fields. We vary $\eta$ between 0.1 and 0.5. Black lines show results for a quadrupole trap ($N=2$), while red and olive data points show results for the hexapole and octupole traps, respectively. In \textbf{a}, the trends of the hexapole and octupole data follow similar shapes to the lines of the quadrupole trap, but with a shallower gradient and an overall lower $E_\text{kin}$. Meanwhile in \textbf{b} we see indications that the optimal value of $\eta$ in an octupole trap is higher than in a quadrupole trap, and gives access to lower energies. The gradients of the MIT trends in \textbf{b} are also shallower than in the quadrupole trap. The lowest temperatures are therefore reached in traps of higher $N$, and the consequences of a sub-optimal choice of $\eta$ are most severe for traps with fewer poles.

To achieve the lowest possible ion energies in a MIT of high $N$ requires well-compensated stray electric fields and a high-amplitude voltage signal. This makes the use of high-order traps for buffer gas cooling practically challenging. When cooled to the $\mu$K range, an ion in the simulated trap will oscillate with an amplitude on the scale of $10^{-4}\times R_\text{rf}$. Since $\eta_j$ scales with $(r_\perp/R_\text{rf}^2)^{(N-2)}$, traps of increasing $N$ require increasing $v_\text{rf}$ to maintain a reasonable stability. For example, an octupole trap with $\eta=0.2$ requires $v_\text{rf}=4.72$\,GV, while the same stability parameter can be achieved in a quadrupole trap with $v_\text{rf}=62.95$\,V. This also has severe implications for axial EMM in the trap, as these voltages are likely to lead to significant rf pickup on the dc electrodes. Even with a suitable choice of $v_\text{rf}$, the $N$-dependent scaling of $\eta$ with $r_\perp$ means that traps with high $N$ are more likely to become unstable when stray electric fields are present. These practical limitations should be considered when selecting how many poles would make a MIT most suitable for buffer gas cooling.

\section{Hybrid optical-Paul trap}\label{sec:hybridtrap}
Another possible extension of the Paul trap is to combine it with an optical tweezer for the ion. This hybrid optical-Paul trap would give an additional confinement for the ion which could benefit the attainable ion kinetic energies and suppress the impact of stray fields. Several approaches of using optical ion trapping in atom-ion setups have been experimentally explored~\cite{Schaetz,Karpa,Sias}, as an optical trap avoids the ion-heating effects present in time-dependent trapping potentials. However, the optical trap for an ion is typically less deep than an rf trap, meaning ions that gain energy due to inelastic collisions are more easily lost. Here, we therefore focus on a hybrid approach, designed to combine the benefits of rf and optical traps. Unlike the traps discussed so far, a deep optical-rf hybrid trap will also affect the density and temperature of the atom cloud. This leads to a tightly-compressed buffer gas and off-resonant heating of the atoms. However, here we focus on the effect on the ion alone, and assume in our simulations that the atoms are unaffected by the optical potential.

\begin{figure}[!t]
\centering
\includegraphics[width=40pc]{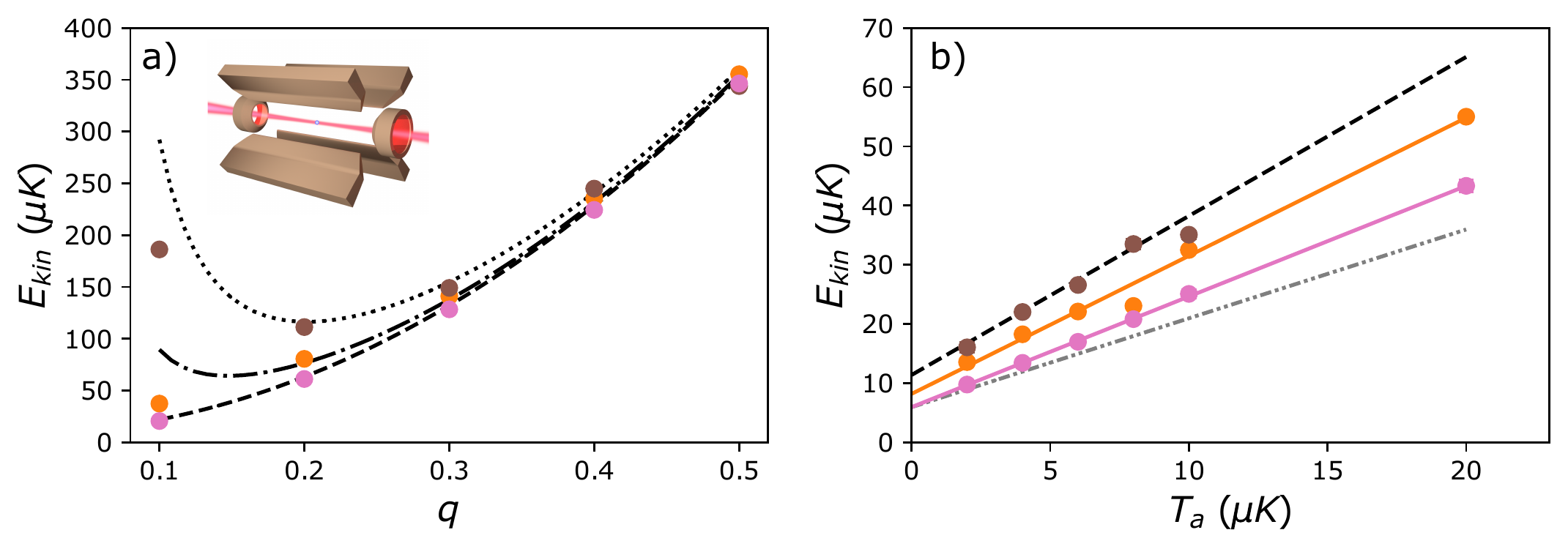}
\caption{Ion energy after buffer gas cooling in an optical-rf hybrid trap. 
The tweezer has a radial frequency $\omega_\text{opt}/\omega_\text{sec}$ of 0.23 (brown), 0.728 (orange) and 2.05 (pink). \textbf{a} Variation with trap parameter q. The black trend lines are simulations for a Paul trap alone with varying radial stray fields of 0.1 (black, dotted), 0.05 (black, dash-dotted) and 0\,Vm$^{-1}$ (black, dashed). In this plot, the stray electric field for the data points is 0.1\,Vm$^{-1}$. Inset: illustration of a hybrid optical-rf trap. \textbf{b} Varying buffer gas temperature. The black, dashed line represents a Paul trap without radial stray fields, while the grey dash-dotted line represents a unity gradient. Data points are taken for a setup with no radial stray field. Errors are standard errors and are sometimes smaller than the point size.}
\label{Tweezer}
\end{figure}

The modeled hybrid optical-Paul trap consists of a quadrupole linear Paul trap with an optical tweezer directed along its axis for additional ion confinement. The setup is shown in the inset of Figure~\ref{Tweezer}. This corresponds to using the rf potential of~\eqref{eq:vtrap_lin} and adding to it a Gaussian potential
\begin{equation}
    V_\text{opt}(r_\perp,r_z)=-U\left(\frac{1}{1+\left(\frac{r_z}{z_\text{R}}\right)^2}\right)\exp\left({-\frac{2r_\perp^2}{w_0^2\left(1+\left(\frac{r_z}{z_\text{R}}\right)^2\right)}}\right)\,.
\end{equation}
Here $r_\perp = \sqrt{r_x^2+r_y^2}$, $w_0$ is the beam waist of our optical tweezer, $z_\text{R}=\frac{\pi w_0^2}{\lambda}$ is the Rayleigh length, and $\lambda$ is the frequency of the tweezer beam. The effective trap depth $U$ is determined by the beam power $P$ and the polarizability $\alpha$ of the species in the trap. It is defined as $U = \frac{\alpha P}{\pi\epsilon_0 c w_0^2}$, with $\epsilon_0$ the permittivity of a vacuum and $c$ the speed of light. The effect of the optical potential on the ion motion is to give an additional position-dependent force towards the trap centre
\begin{equation}
    F_{\text{opt},j} = -\frac{\text{d}V_\text{opt}}{\text{d}r_j} = -m_\text{i}\omega_{\text{opt,}j}^2 r_{j}\,.
\end{equation}
Here $\omega_{\text{opt,}j}$ is the trapping frequency of the optical trap in each direction ${\omega_{\text{opt,}x,y}=\sqrt{\frac{4U}{m_\text{i}w_0^2}}}$, $ \omega_{\text{opt,}z}=\sqrt{\frac{2U\lambda^2}{m_\text{i}\pi^2w_0^4}}$~\cite{Grimmreview}. The confinement of the tweezer in the axial direction is sufficiently weak for us to assume that $\omega_{\text{opt,}z}$ plays no significant role here.

We simulate buffer gas cooling of an ion in this hybrid optical-Paul trap, as a function of $T_\text{a}$ and $q$ for varying optical confinements. The contribution from the linear Paul trap is taken with identical trap parameters and geometry to those used in section~\ref{sec:lin}. We choose a stray radial electric field of either 0 or 0.1\,Vm$^{-1}$. For the former, no other EMM effects are included. The optical tweezer has a wavelength of 1070\,nm, a variable power set to 0.423, 4.23 and 2.12\,W and a variable beam waist set to 2, 2 and 1\,$\mu$m, respectively. The Stark shift of a $^{171}$Yb$^+$ ion in such a tweezer gives it a dynamic polarizability $\alpha$ of approximately 66\,au (1.09$\times 10^{-39}$\,Jm$^2$V$^{-2}$)~\cite{Yb+polarizability}. The chosen tweezer parameters therefore correspond to respective trapping frequencies $\omega_{\text{opt,}x,y}/\omega_{\text{sec,}x,y}=0.230, 0.728$ and $2.05$ for the $^{171}$Yb$^+$ ion. Here, trapping frequencies are given as a ratio of the linear Paul trap's secular frequency, as defined in section~\ref{sec:lin}.

Our simulations show that adding an optical tweezer aids in decreasing the effects of radial stray fields on the ion cooling. Therefore at low $q$, the ion can be cooled to temperatures significantly lower than those accessible in a Paul trap alone. Figure~\ref{Tweezer}\textbf{a} shows the simulation data of $E_\text{kin}$ for variable $q$ in a hybrid trap with stray electric fields of 0.1Vm$^{-1}$. The data is compared to the trend lines from a pure Paul trap with varying levels of stray fields. All lines and data points in this plot also include $E_\text{ax}=42.4$\,$\mu$K and $E_\text{ph}=9.18$\,$\mu$K at $q=0.219$ and $\Omega_\text{rf}=2\pi\times 2$\,MHz. For a hybrid trap with increasing $\omega_\text{opt}$, the data points tend increasingly towards the trend line of an rf trap with no stray electric fields. In other words, the tighter the optical tweezer, the smaller the heating effects of radial EMM, allowing the ion to reach much lower temperatures than in an equivalent regular Paul trap. The same suppression of EMM effects is seen when analysing simulations of a hybrid trap with variable $\Omega_\text{rf}$.

A tighter confinement of the optical tweezers leads to a more efficient buffer gas cooling process. Figure~\ref{Tweezer}\textbf{b} shows the linear relationship between the ion energies reached in a hybrid trap of variable optical depth, and the temperature $T_\text{a}$ of the buffer gas in which it is submerged. We also plot the same relationship for a Paul trap without an optical tweezer as well as a line with unity gradient, for comparison. The latter corresponds to a static trap. We find that the hybrid trap results follow lines of increasingly shallow gradient for an increasingly tight optical tweezer. This reflects the time-independent tweezer potential becoming increasingly dominant over rf contributions in determining the energy reached by the ion. The value of $E_\text{kin,i}$ at $T_\text{a}=0$\,K also decreases for an increasingly tight tweezer.

\begin{figure}[!t]
\centering
\includegraphics[width=20pc]{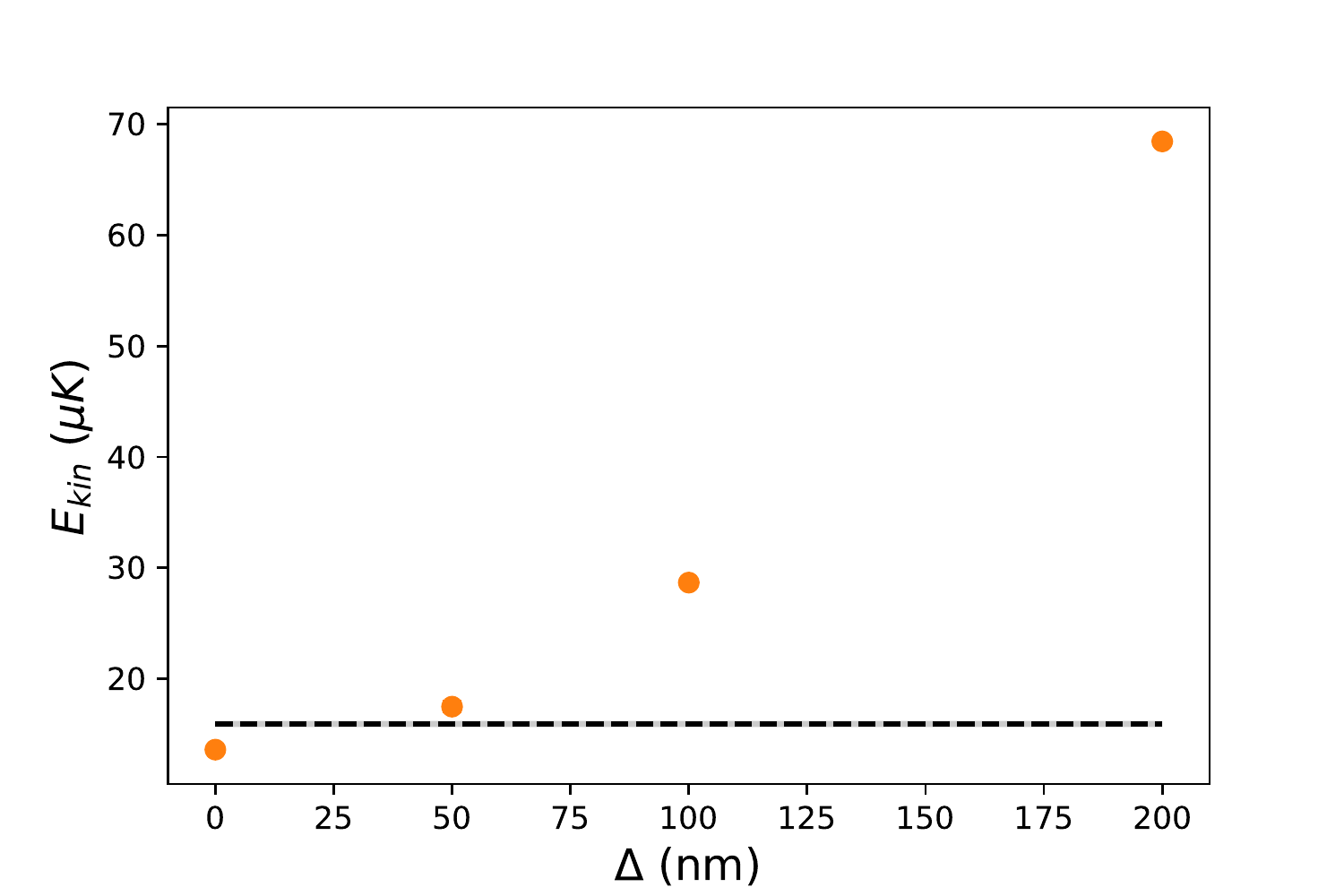}
\caption{Ion energy after buffer gas cooling in a displaced rf-optical hybrid trap. Orange data points mark the temperatures reached for a variable radial offset ($\Delta$) between the Paul trap axis and the optical beam. The black dashed line shows the energy achieved when simulating the same process with the dipole beam switched off. Errors are standard errors and are sometimes smaller than the point size.}
\label{Tweezer Offset}
\end{figure}

To benefit from the optical tweezer, it must be perfectly aligned with the Paul trap axis. Figure~\ref{Tweezer Offset} shows the simulation of buffer gas cooling in a hybrid trap in which the optical tweezer runs parallel to the Paul trap axis, with a varying radial offset $\Delta$ between them. For comparison, we also indicate the ion energy reached in an equivalent Paul trap with no optical tweezer. Offsets of  50\,nm or greater lead to ion energies above this value, and the greater the offset, the higher the ion energy. For misalignment, the effect of the tweezer is to pull the ion from the Paul trap axis, providing an additional source of excess micromotion, akin to a radial stray field.

Another important limitation for practical implementation of the optical-rf hybrid trap is the effect of the optical potential on the atoms. At the high trapping frequencies required to observe a significant benefit of the tweezer on the ion, heating and confinement of the atoms become significant. If we wish to trap an $\mathrm{Yb}^+$ ion in a tweezer with $\lambda=1070$\,nm and $\omega_{\text{opt,}x,y}=2.05\times\omega_{\text{sec,}x,y}$, a  cloud of Li atoms in the path of the beam will experience a trap depth of 82\,mK~\cite{Grimmreview}. This will lead to a tightly-compressed atom cloud. Combined with the high photon-intensity of the tweezer, these atom densities give a possibility for unwanted chemistry~\cite{TomzaPhotoa,Mohammadi2021LDC}. The same tweezer would also lead to heating of the Li atoms due to photon scattering at a rate of 128\,$\mu$Ks$^{-1}$, significantly reducing the efficiency of buffer gas cooling. To address the problem of trap depth, it may be possible that for some atom/ion species combinations one could confine the ion in an optical tweezer with a wavelength equal to a tune-out wavelength of the atoms, as has been proposed for multi-species atom experiments~\cite{LeBlanc2007SSO}. However for our Li/Yb$^+$ species, this is not possible, as the only Li tune-out wavelength that is red-detuned for Yb$^+$ is too close to the D1 and D2 lines of Li~\cite{Lituneout}. Suitable combinations of bichromatic optical dipole traps can also be used for all-optical trapping of both atoms and ions~\cite{arxivKarpa}. We consider this as an alternative method of reducing the effect of the tweezer on the atom cloud in an optical-rf hybrid trap. However, the cooling in a bichromatic trap is highly sensitive to relative position fluctuations between the two trapping beams. Furthermore, for our species and desired trap depth, the wavelength and intensity requirements cannot be met simultaneously with the requirement for a large detuning of each beam from the atom and ion transition lines, to avoid off-resonant heating. In fact, the heating problem would be intensified by the higher-frequency beam.

\begin{figure}[!t]
\centering
\includegraphics[width=20pc]{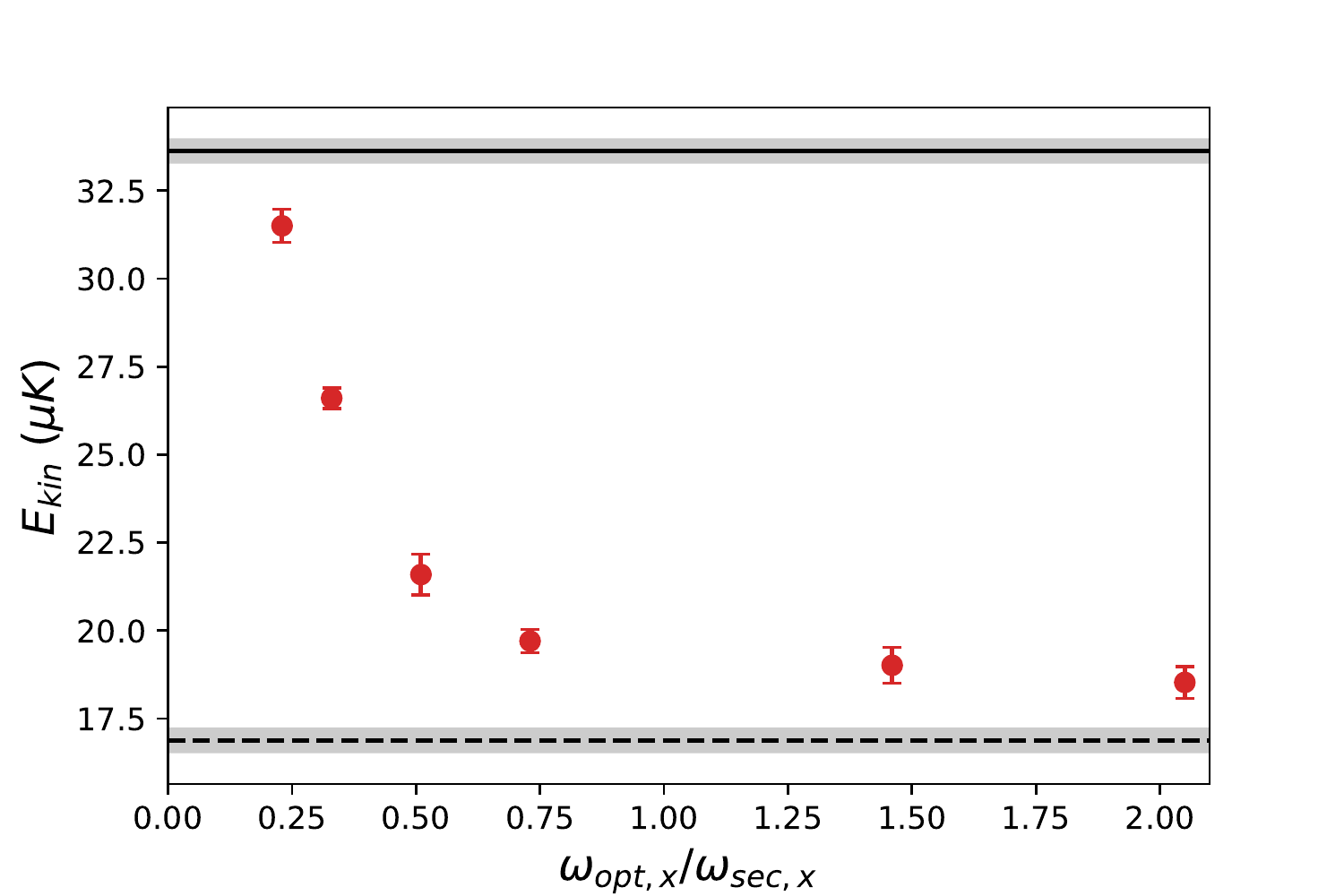}
\caption{Energy of one ion in a two-ion crystal after buffer gas cooling, with the second ion confined by an optical tweezer outside of the atom  bath. Data points show results for a tweezer of varying depth. For comparison, the black lines show the temperature reached by an ion in a Paul trap without a tweezer for the single ion case (black, dashed) and two-ion case (black, solid). Errors are standard errors. The grey shaded regions indicate standard errors on the lines.}
\label{2ion Tweezer}
\end{figure}

A way to avoid the negative effects of the tweezer on the atoms may be to use a two-ion crystal, with one ion submerged in the atomic bath and the other in the tweezer. Using the ion chain's collective modes of oscillation, the pinned ion could then be used to sympathetically enhance the cooling of the ion in the buffer gas. We simulate this option and show the results in figure~\ref{2ion Tweezer}. It shows the energy reached by the untweezed ion in the crystal, dependent on the trapping frequency of the optical tweezer. We also show for comparison $E_\text{kin}$ for a single ion (dashed line) and a two-ion crystal (solid line) after buffer gas cooling in a Paul trap without an optical tweezer. The results show that the higher the trapping frequency of the optical tweezer, the lower the achievable ion energies. However, these energies tend towards a minimum equal to the energy of a single ion in a Paul trap, for large tweezer depths. The effect of the tweezer on the two-ion crystal is to decouple the motion of the two ions, thereby giving no additional cooling benefits.

\section{The quantum dynamics of micromotion-induced heating}\label{sec:quantum}

Up until now, we have used classical physics to describe the dynamics of the trapped ion in the buffer gas. However, at the simulated temperatures, quantum effects may play a role. In particular, the angular momentum in the atom-ion collision and the energy of the ion in its trap will both be quantized. Furthermore, the dispersion of the atomic wavepacket may become much larger than the micromotion amplitude of the ion such that it cannot be resolved at all in a collision. 

Specifically, when looking into micromotion-induced heating due to the time-dependence of the trap, the classical and quantum mechanical picture might differ. In the classical simulations of colliding atoms and trapped ions, the heating stems from a temporary binding between the ion and atom during which multiple collisions can occur~\cite{Cetina}. At some point, and depending on the phase of the rf trap, the ion and atom can violently separate again. During the whole process, energy can be absorbed from the rf field, leading to heating of the system. One may wonder whether this behaviour survives in a quantum treatment, in which binding can only occur in  discrete molecular states with energy level spacings that may be larger than the collision energy. In this section we answer the question: Can quantum mechanics save us from micromotion-induced heating in an atom-ion collision?  

One approach is to consider the same case that we studied in the classical regime, but this time launching an atomic wavepacket at an ionic wavepacket that is held in a (quantized) rf Paul trap.  Floquet theory allows us to solve the Schr\"{o}dinger equation, similar to previous theoretical investigations into mechanically-controlled interactions between trapped atoms and ions~\cite{Idziaszek2007CCS,Idziaszek2011MQD,Nguyen2012MMT,Joger2014QDA,Gerritsma2012BJJ} and individually-trapped atoms~\cite{Krych2009CCT}. However, even if we restrict ourselves to one  dimension to reduce the complexity, the required Floquet Hilbert space is very large and convergence is very slow. Instead, we follow a qualitative analysis of the problem, showing that quantum mechanics cannot eliminate micromotion-induced heating.

\subsection{Quantum description of an ion in a Paul trap}
\label{subsec:qion}
We start by studying the ion motion quantum mechanically, when the ion is confined by an rf-potential $\hat V_{\rm rf}(t)=\frac{1}{8} m_\text{i} \Omega_{\rm rf} \left(a+2q\cos\left(\Omega_{\rm rf} t\right )\right) \hat x_\text{i}^2$. Its motion can be described by the Hamiltonian $\hat H_{\rm i}(t)=\hat p_\text{i}^2/2m_\text{i} + \hat V_{\rm rf}(t)$, with wavefunction $\vert \Psi_\text{i}(t) \rangle$ and momentum $\hat p_\text{i}$. Using the Cook and Shankland transformation the wavefunction of the ion can be written as $\vert \Psi_{\rm CS} (t)\rangle  = \exp\left[-i/(4\hbar) m_\text{i} q \Omega_{\rm rf} \hat{x}^2 \sin\left(\Omega_{\rm rf} t\right)\right]\vert \Psi_\text{i}(t) \rangle$~\cite{Cook1985QTP}. The fast dynamics is now contained in the exponential factor and the slow-varying part of the ion wavefunction evolves under the Hamiltonian $\hat H_{\rm i}(t)$. The rf-potential $\hat V_{\rm rf}(t)$ can be separated into a secular and intrinsic part by 
\begin{equation}
\hat V_{\rm rf}(t)= \frac{1}{2}m_\text{i}\omega_\text{sec,i}^2\hat x_\text{i}^2+\hat H_\text{mm}(t)\,,
\end{equation}
where $\omega_\text{sec,i}$ takes the same value as in our classical equations (see section~\ref{sec:lin}). Here, the ion intrinsic micromotion Hamiltonian $\hat H_\text{mm} (t)$ is given by 
\begin{equation}
\hat H_\text{mm}(t)=-m_\text{i}g^2\omega_\text{sec,i}^2\hat x_\text{i}^2\cos(2\Omega t)-g\omega_\text{sec,i} \{\hat x_\text{i},\hat p_\text{i}\}\sin(\Omega_{\rm rf} t)\,.
\end{equation}
Furthermore, $g=[2(1+2a/q^2)]^{-1/2}$ and $\{.,.\}$ denotes the anti-commutator. Note that excess micromotion is not included in our quantum study.

Typically in analysis of the trapped ion, one would proceed with applying the secular approximation. The basis for the secular approximation is to neglect $\hat H_\text{mm}(t)$ such that the dynamics of the ion is given by a static harmonic oscillator alone. Note that the relative size of the terms in $\hat H_\text{mm}(t)$ are of a similar size to those of the static harmonic oscillator and in fact, the mathematical basis for the secular approximation is quite delicate. The reason it works at all is the structure of the eigenstates of the harmonic potential. The time-dependent part $\hat H_\text{mm}(t)$ contains terms $\propto  \hat{x}_\text{i}^2$. Taking the Fock states of the static harmonic oscillator as a convenient basis in which to study the Hamiltonian $\hat H_{\rm i}(t)$, we see that strict selection rules apply, i.e. $\langle n|\hat{x}_\text{i}^2|m\rangle=0$ except for $n=m\pm 2$ and $n=m$, with $|n\rangle$ denoting the $n$-th Fock state. Since $\Omega_\text{rf}\gg \omega_\text{sec,i}$ such transitions cannot be resonant. Therefore, the only effect of $\hat H_\text{mm}(t)$ here is to cause a small increase in the effective secular trapfrequency that becomes more prominent for large $q$. 

However, any anharmonicity in the trapping potential can cause a serious breakdown of the secular approximation. This is because transitions $|n\rangle\rightarrow |m\rangle$ close to the resonance condition $|\omega_\text{sec,i}(n-m)|=\Omega_\text{rf}$ are allowed in anharmonic traps. The observation of such parametric resonances in rf traps, in which ions show rapid heating,  dates back a long time~\cite{Alheit1995OIP,Pedregosa2010ACR,DengDBS2015}. Thus, for anharmonic potentials, the contribution of $\hat H_\text{mm}(t)$ can not be ignored. A similar effect is expected when describing an ion in the presence of a buffer gas, which adds an additional atom-ion interaction potential.  

\subsection{Quantum description of atom-ion interactions}

Now we add the atom to our quantum description. Then, the Hamiltonian is given by
\begin{equation}
\label{eq:Htotalq} 
    \hat H(t)=\hat H_\text{a}+\hat H_\text{i}(t)+\hat V_\text{ai}\\
    =\frac{\hat p_\text{a}^2}{2m_\text{a}}+\frac{\hat p_\text{i}^2}{2m_\text{i}}+\hat V_\text{ai}+\hat V_{\rm rf}(t)\,,
\end{equation}
where $\hat V_{\rm ai}=-C_4/\hat r^4 $ is the atom-ion interaction potential, $r$ is the separation between the atom and the ion and $C_4$ takes the same value as that in our classical simulations (see section~\ref{sec:method}). We transform to the  relative ($r$) and centre of mass ($R$) coordinates, to obtain the atom-ion dynamics~\cite{Joger2014QDA}. In these coordinates, eq.~\eqref{eq:Htotalq} can be written as
\begin{align} 
\nonumber \hat H(t)&=\hat H_\text{R}(t)+\hat H_\text{r}(t)+\hat H_{Rr}(t)\,, \\
\label{eq:qHR0}
    \hat H_R(t)&=-\frac{\hbar^2}{2M}\frac{\partial^2}{\partial R^2}+\frac{1}{2}M\omega_R^2\hat R^2+\hat H^{RR}_\text{mm}(t)\,,\\
\label{eq:qHr0}
    \hat H_r(t)&=-\frac{\hbar^2}{2\mu}\frac{\partial^2}{\partial r^2}+\frac{1}{2} \mu\omega_r^2\hat r^2 -\frac{C_4}{\hat r^4}+\hat H^{rr}_\text{mm}(t)\,,\\
\label{eq:qHRr1}
    \hat H_{Rr}(t)&=\mu\omega_\text{sec,i}^2\hat R\hat r+\hat H^{Rr}_\text{mm}(t)\,,
\end{align}
\noindent with $\omega_R=\omega_\text{sec,i}\sqrt{m_\text{i}/M}$ and $\omega_r=\omega_\text{sec,i}\sqrt{
m_\text{a}/M}$ denoting the centre of mass and relative frequencies of the system, respectively. The micromotion Hamiltonians are  
\begin{align}
   \hat  H^{RR}_\text{mm}(t)&=-Mg^2\omega_R^2\hat{R}^2\cos(2\Omega t)-g\omega_R\{\hat{R},\hat{P}\}\sin(\Omega_\text{rf}t)\,,\\
    \hat H^{rr}_\text{mm}(t)&=-\mu g^2\omega_r^2\hat{r}^2\cos(2\Omega t)-g\omega_r\{\hat{r},\hat{p}\}\sin(\Omega_\text{rf}t)\,,\\
    \hat H^{Rr}_\text{mm}(t)&=-\mu g^2\omega_\text{sec,i}^2\hat{r}\hat{R}\cos(2\Omega t)-g\omega_\text{sec,i}\left(\frac{\mu}{M}\{\hat{r},\hat{P}\}+\{\hat{R},\hat{p}\}\right)\sin(\Omega_\text{rf}t)\,.
\end{align}
Here $\hat P$ and $\hat p$ are the centre of mass and relative momenta, respectively.

To understand where heating can arise in this system, we must look at each of the above equations individually. Let us begin with the centre of mass Hamiltonian  eq.~\eqref{eq:qHR0}. Excluding micromotion effects (i.e. $\hat H^{RR}_\text{mm}(t)=0$), the centre of mass motion can be treated as that of a harmonic oscillator, with solutions that are Fock states $|n_R\rangle$ and energies $n_R\hbar\omega_R$. If we include micromotion, $\hat H_R(t)$ resembles the Hamiltonian $\hat H_\text{i}(t)$ of a single ion in a Paul trap, as described in section~\ref{subsec:qion}, now with $m_\text{i}\rightarrow M$ and $\omega_\text{sec,i}\rightarrow\omega_R$. Since this Hamiltonian is without anharmonicity, it describes a stably trapped particle for which we may employ the secular approximation. As a result, looking at the centre of mass motion alone, it would seem that no significant heating is expected.

The relative Hamiltonian, eq.~\eqref{eq:qHr0}, takes a similar form to that of the centre of mass, except it includes the additional static potential $\hat V_\text{ai}$. Provided that there is no micromotion (i.e. $\hat H^{rr}_\text{mm}(t)=0$), this static term does not give rise to heating effects. However, it does mean that, unlike the centre of mass Hamiltonian, the solutions to the unperturbed relative Hamiltonian are not Fock states. Instead the dynamics must be obtained by solving the Schr\"odinger equation by means of quantum defect theory~\cite{Joger2014QDA}. In the limit where $r\rightarrow 0$, the energy is dominated by the atom-ion interaction, and we obtain even and odd solutions given by~\cite{Idziaszek2007CCS,Idziaszek2011MQD}

\begin{eqnarray}\label{eq_asymp_a}
\tilde{\psi}_e(r)&\propto&|r|\sin \left(\frac{R^*}{|r|}+\phi_e\right)\,,\\
\label{eq_asymp_b}
\tilde{\psi}_o(r)&\propto& r \sin \left(\frac{R^*}{|r|}+\phi_o\right)\,,
\end{eqnarray}

\noindent where we set the even and odd short range phases as $\phi_e=\pi/4$ and $\phi_o=-\pi/4$. We can use Eqs.~\eqref{eq_asymp_a},\eqref{eq_asymp_b} as boundary conditions in a Numerov method to find the relative eigenstates $|r\rangle$ and energies $\epsilon_r$ of~\eqref{eq:qHr0}. The eigenstates can be subdivided in two types. On the one hand we find many states that are mostly confined by the harmonic part of~\eqref{eq:qHr0} and we find a number of states that are deeply bound by the $\hat V_\text{ai}$ term, which we identify as molecular states. Note that the basis states of the total unperturbed Hamiltonian~\eqref{eq:Htotalq} ($\hat H_\text{mm}(t)=0$) are therefore given by $|\Phi_{rn}^\text{(b)}\rangle=|r\rangle\otimes |n_R\rangle$.

When including micromotion in eq.~\eqref{eq:qHr0}, we again see similarities to eq.~\eqref{eq:qHR0}, except for the addition of $\hat{V}_\text{ai}$. The Hamiltonian for the relative coordinate resembles that of a single ion in a Paul trap with $m_\text{i}\rightarrow \mu$ and $\omega_\text{sec,i}\rightarrow\omega_r$, but with $\hat{V}_\text{ai}$ acting as an additional anharmonic term. This anharmonicity becomes a problem now that we include micromotion, as the time-dependent terms will cause parametric heating in analogy to heating observed in anharmonic rf traps~\cite{Alheit1995OIP,Pedregosa2010ACR,DengDBS2015}. Note that due to the dense spectrum of nearly-free scattering states, there will always be (near) resonant transitions available for the parametric heating.

Finally, we consider the role of eq.~\eqref{eq:qHRr1}, which includes contributions from both the centre of mass and relative motions. Its contribution to the total Hamiltonian~\eqref{eq:Htotalq} can be interpreted as a further anharmonicity, in addition to $\hat V_\text{ai}$, with implications for both the relative and centre of mass dynamics. This anharmonicity is not a problem until we include non-zero micromotion terms, as the combination of micromotion and an anharmonic potential gives rise to allowed transitions close to a resonance. In the case of the total Hamiltonian~\eqref{eq:Htotalq}, the crossterm $\hat H^{Rr}_\text{mm}(t)$ drives transitions $|n_R\rangle|r\rangle \rightarrow |n_R+ 1\rangle|r'\rangle$ that follow the resonance condition $|\epsilon_r-\epsilon_r'|=\hbar\omega_\text{i}+k \hbar\Omega$ with $k=0,\pm 1, \pm 2$. This leads to heating in both the relative and centre of mass coordinates. Physically, it can be interpreted as the system absorbing and emitting quanta of energy $\hbar\Omega$ from the trap driving. By investigating the roles of $\hat H_{Rr}(t)$ and $\hat V_\text{ai}$ in this way, it becomes clear that quantum mechanics is unlikely to save us from micromotion-induced heating.

\begin{figure}[!t]
\centering
\includegraphics[width=20pc]{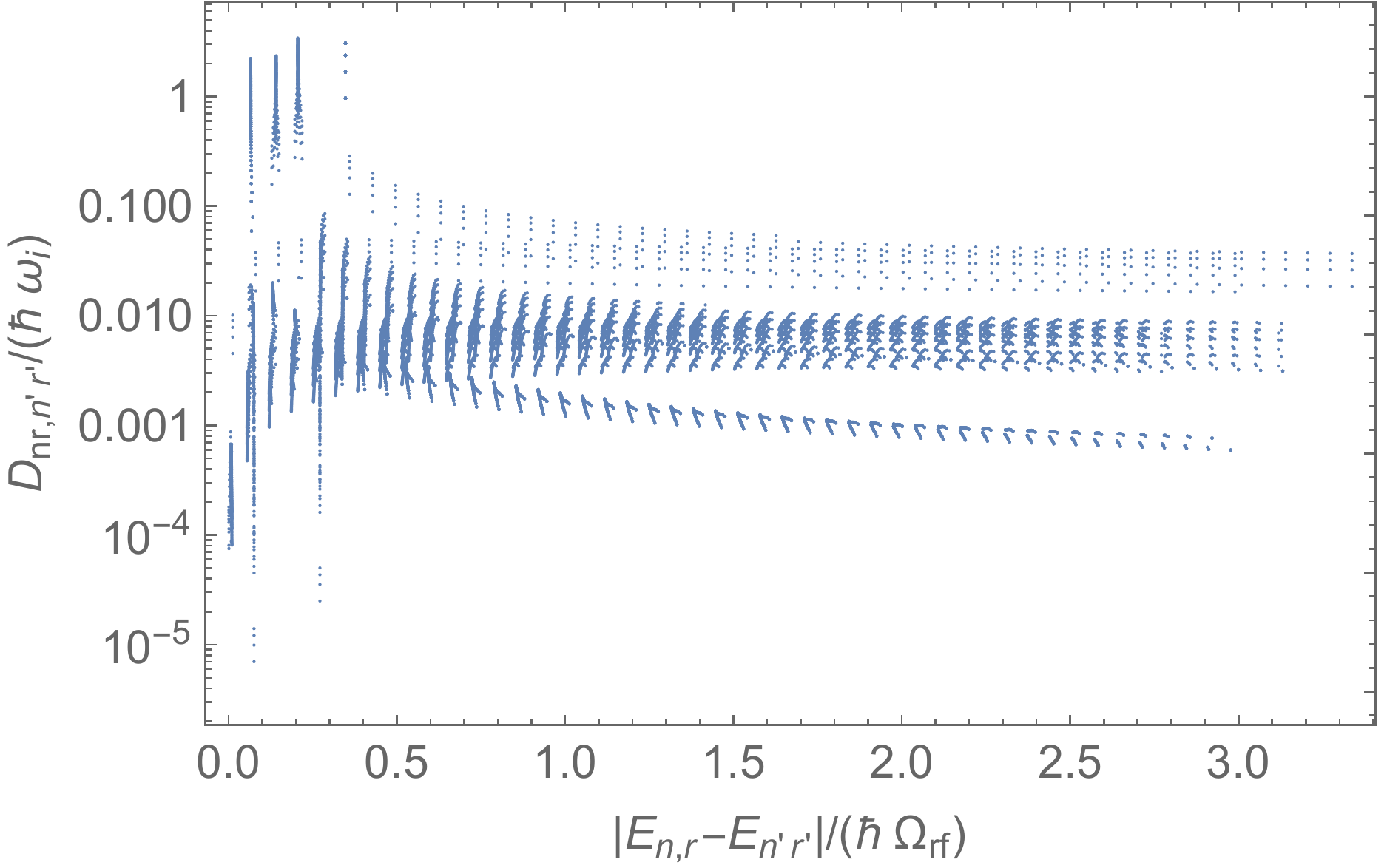}
\caption{Matrixelements $D_{nr,n'r'}=|\langle n_R|\langle r|H_\text{mm,1}|r'\rangle|n'_R\rangle |$ versus the energy difference $\delta E=|E_{nr}-E_{n'r'}|$. Data is shown for a system of $^6$Li and $^{171}$Yb$^+$, confined in a Paul trap with parameters $\Omega_\text{rf}=2\pi\times 2$\,MHz, $a=0$ and $q=0.5$. } 
\label{MM_matelems}
\end{figure}

Having established the sources of anharmonicity in the system, we now further investigate the micromotion terms of~\eqref{eq:Htotalq}, which perturb the dynamics of the system and combine with the anharmonic potential to give rise to parametric heating. In particular, we will investigate the expected scale of this micromotion perturbation, by looking at the full hamiltonian and calculating the matrix elements $\langle n_R|\langle r| H_\text{mm,1}^\text{max}|r\rangle|n_R\rangle$. Here $H_\text{mm,1}$ denotes the prefactor to the terms $\propto\sin\Omega_\text{rf}t$  that dominate the micromotion Hamiltonians $\hat H^{RR}_\text{mm}(t)$, $\hat H^{rr}_\text{mm}(t)$ and $\hat H^{Rr}_\text{mm}(t)$. ~\cite{Joger2014QDA}.   We consider the realistic scenario where $\Omega_\text{rf}=2\pi\times 2$\,MHz, $a=0$ and $q=0.5$, similar to the experimental parameters of ref.~\cite{Nature}. The results are shown in Fig.~\ref{MM_matelems}. 

We see that the couplings due to micromotion can be up to $\sim 0.1\hbar\omega_\text{sec,i}$ for an energy difference $\delta E \sim \hbar\Omega_\text{rf}$. This energy scale is comparable to the fractional change of kinetic energy $K$ for a classical hardcore collision. Assuming the atom is at rest initially and the initial ion kinetic energy is $K_1$, we get for the final ion kinetic energy $K_2=\left(1-\frac{2\beta}{(1+\beta)^2}\right)K_1=0.93K_1$, with $\beta=m_\text{a}/m_\text{i}$. For a cold ion $K_1\sim\hbar\omega_\text{sec,i}$, giving $K_1-K_2\sim 0.07 \hbar\omega_\text{sec,i}$. We therefore conclude that buffer gas cooling to the ion ground state is not possible, even in the quantum picture, as micromotion-induced heating will impose a limit on the ion temperature, similar in scale to that seen in classical analysis.

\section{Conclusion and Outlook}
In this work, we demonstrate that careful optimization of ion trap parameters is crucial for reaching cold temperatures when buffer gas cooling ions with ultracold atoms in time-dependent traps. We show that, under appropriate conditions, buffer gas cooling should allow one to reach the regime of quantized ion motion. For the Li/Yb$^+$ system, parameter optimization can allow buffer gas cooling to ion kinetic energies three times colder than the previous best experimental values, corresponding to collision energies approximately 0.6 times the $s$-wave limit. This makes buffer gas cooling capable of  outperforming laser Doppler cooling in some situations, and it is even competitive when compared to some sub-Doppler cooling techniques such as Sishyphus cooling. These low energies are achievable by suppression of micromotion-induced heating effects, through optimization of $q$ and $\Omega_\text{rf}$, and by compensation of trap imperfections to low, but experimentally realistic, levels. Additionally, we benefit from a large ion-atom mass ratio. For equal mass systems, similar trap parameter dependencies were found for a linear Paul trap~\cite{Pinkas2020EIT}, however the achievable temperatures for such systems remain far from the $s$-wave regime.

We further demonstrate that the lowest possible energies achievable in a linear Paul trap can also be reached through similar parameter optimization in a range of rf traps. In the case of the multipole and hybrid optical-rf traps, even lower temperatures can be reached, albeit with the requirement for challenging experimental parameters and, for the hybrid trap, a method of avoiding photon-scattering heating effects. In addition, we provide a quantum description of the atom-ion interaction, and argue that micromotion-induced heating continues to play a significant role in this regime.

Our study shows a route towards reaching deeper into the quantum regime with time-dependent traps, and presents buffer gas cooling as a feasible alternative to laser cooling techniques. Especially for studies of quantum chemistry and precision spectroscopy, cold controlled collisions in hybrid atom-ion systems are important. Moreover, in the quantum regime one can benefit from the tunability of the atom-ion interaction through Feshbach resonances~\cite{Idziaszek2011MCQ,Tomza2015CIY,SchaetzFRs} and use the ground state cooled-ion for high-precision studies of charged impurity physics~\cite{Astrak2021IPBEC, Oghittu2021doa}. 

\section{Acknowledgements}
This work was supported by the Netherlands Organization for Scientific Research (Vidi Grant 680-47-538 and Start-up grant 740.018.008, and Vrije Programma 680.92.18.05 (R.G.)).  R.S.L. acknowledges funding from the European Unions Horizon 2020 research and innovation programme under the Marie Sklodowska-Curie grant agreement No 895473. A.S.N is supported 
by the Dutch Research Council (NWO/OCW), as part of the Quantum Software Consortium programme (project number
024.003.037).

\section{Appendix A}
\begin{table}[hbt!]
\centering
\begin{tabular}{ |c|c|c|} 
 \hline
 Fit parameter & Trend in $q$ & Trend in $\Omega_\text{rf}$\\
 \hline
 $A_x$ & 2.483$\times 10^{-8}$ & 1.335$\times 10^{-8}$\\
 $A_y$ & 2.488$\times 10^{-8}$ & 1.335$\times 10^{-8}$\\
 $A_z$ & 1.859$\times 10^{-7}$ & 1.335$\times 10^{-7}$\\
 $b$ & 1.782 & 1.690\\
 $c$ & 1.689 & 1.704\\
 $d$ & 1.688 & 1.771\\
 $g$ & 0 & 9.871\\
 \hline
 \end{tabular}
 \caption{Fit parameters for the ion energy and motion after buffer gas cooling in a linear Paul trap. These parameters together with equation~\eqref{eq:totalKEfit} are used to obtain the lines in Fig~\ref{fig:qdrlinPaul}\textbf{a} and \textbf{b} for varying radial micromotion contributions. Note that these results were obtained not only by fitting to the data shown in figure~\ref{fig:qdrlinPaul}, but also to data from similar simulations that excluded one or more forms of EMM.}
 \end{table}


\section{References}
\begin{thebibliography}{}
\bibitem{Tomzareview} M. Tomza, K. Jachymski, R. Gerritsma, A. Negretti, T. Calarco, Z. Idziaszek, and P. S. Julienne, Rev. Mod. Phys. \textbf{91}, 035001 (2019).

\bibitem{Harter2014CAIrev} A. H\"{a}rter and J. H. Denschlag, Contemp. Phys. \textbf{55}, 33 (2014).
\bibitem{Cote2016UHAI} R. Côté, Adv. Atom. Mol. Opt. Phys. \textbf{65}, 67 (2016).

\bibitem{Ratschbacher2013DSI}L. Ratschbacher, C. Sias, L. Carcagni, J. M. Silver, C. Zipkes, and M. Köhl, Phys. Rev. Lett. \textbf{110}, 160402 (2013).
\bibitem{Schmid2010DCT}S. Schmid, A. H\"{a}rter, and J. H. Denschlag, Phys. Rev. Lett. \textbf{105}, 133202 (2010).
\bibitem{Furst2018DSI}H. F\"{u}rst, T. Feldker, N. V. Ewald, J. Joger, M. Tomza, and R. Gerritsma, Phys. Rev. A \textbf{98}, 012713 (2018).
\bibitem{Kollath2007STM}C. Kollath, M. Köhl, and T. Giamarchi, Phys. Rev. A \textbf{76}, 063602 (2007).
\bibitem{Astrak2021IPBEC}G. E. Astrakharchik, L. A. Peña Ardila, R. Schmidt, K. Jachymski and A. Negretti, Communications Physics \textbf{4}, 94 (2021).
\bibitem{Dieterle2021TSI}T. Dieterle, M. Berngruber, C. Hölzl, R. Löw, K. Jachymski, T. Pfau, and F. Meinert, Phys. Rev. Lett. \textbf{126}, 033401 (2021).
\bibitem{Hirzler2020CNC}H. Hirzler, E. Trimby, R. S. Lous, G. C. Groenenboom, R. Gerritsma, and J. Pérez-Ríos, Phys. Rev. Research \textbf{2}, 033232 (2020).
\bibitem{Zipkes2010TSI}C. Zipkes, S. Palzer, C. Sias and M. K\"{o}hl 
Nature \textbf{464}, 388–391 (2010).
\bibitem{Oghittu2021doa} L. Oghittu, M. Johannsen, R. Gerritsma, A. Negretti, ArXiv 2109.03143 (2021).

\bibitem{Ratschbacher2012CCR} L. Ratschbacher, C. Zipkes, C. Sias and M. Köhl, Nature Physics \textbf{8}, 649–652 (2012).
\bibitem{Haze2015CEC} S. Haze, R. Saito, M. Fujinaga, and T. Mukaiyama, Phys. Rev. A \textbf{91}, 032709 (2015).
\bibitem{Sikorsky2018SCA}T. Sikorsky, Z. Meir, R. Ben-shlomi, N. Akerman and R. Ozeri, Nature Communications \textbf{9}, 920 (2018).
\bibitem{Mohammadi2021LDC} A. Mohammadi, A. Krükow, A. Mahdian, M. Deiß, J. Pérez-Ríos, H. da Silva, Jr., M. Raoult, O. Dulieu, and J. Hecker Denschlag, Phys. Rev. Research \textbf{3}, 013196 (2021).
\bibitem{Bissbort2013ESS}U. Bissbort, D. Cocks, A. Negretti, Z. Idziaszek, T. Calarco, F. Schmidt-Kaler, W. Hofstetter, and R. Gerritsma, Phys. Rev. Lett. \textbf{111}, 080501 (2013).

\bibitem{Doerk2010AIQ} H. Doerk, Z. Idziaszek, and T. Calarco, Phys. Rev. A \textbf{81}, 012708 (2010). 
\bibitem{Sauder1968TEC}W. C. Sauder, J. Res. Natl. Bur. Stand. A. Phys. Chem. \textbf{72A}, 1 (1968).
\bibitem{Chen2014NGS}K. Chen, S. T. Sullivan, and E. R. Hudson, Phys. Rev. Lett. \textbf{112}, 143009 (2014).
\bibitem{Zipkes2011KST}C. Zipkes, L. Ratschbacher, C. Sias and M. K\"{o}hl, New J. Phys. 13 053020 (2011).
\bibitem{Meir2016DGS}Z. Meir, T. Sikorsky, R. Ben-shlomi, N. Akerman, Y. Dallal, and R. Ozeri,
Phys. Rev. Lett. \textbf{117}, 243401 (2016). 


\bibitem{Morigi2000GSLC}G. Morigi, J. Eschner, and C. H. Keitel, Phys. Rev. Lett. \textbf{85}, 4458 (2000).
\bibitem{Ioncooling} W. M. Itano, J. C. Bergquist, J. J. Bollinger and D. J. Wineland, Phys. Scr. \textbf{1995}, 106 (1995).


\bibitem{Idziaszek2011MCQ}Z. Idziaszek, A. Simoni, T. Calarco and P. S. Julienne, New J. Phys. \textbf{13}, 083005 (2011).
\bibitem{Tomza2015CIY}M. Tomza, C. P. Koch, and R. Moszynski, Phys. Rev. A \textbf{91}, 042706 (2015).

\bibitem{Chin2010FR}C. Chin, R. Grimm, P. Julienne, and E. Tiesinga, Rev. Mod. Phys. \textbf{82}, 1225 (2010).
\bibitem{SchaetzFRs}P. Weckesser, F. Thielemann, D. Wiater, A. Wojciechowska, L. Karpa, K. Jachymski, M. Tomza, T. Walker and T. Schaetz, arXiv:2105.09382
\bibitem{Nature} T. Feldker, H. Fürst, H. Hirzler, N. V. Ewald, M. Mazzanti, D. Wiater, M. Tomza and R. Gerritsma, Nature Physics \textbf{16}, 413–416 (2020).
\bibitem{Cetina} M. Cetina, A. T. Grier, and V. Vuleti\'{c},
Phys. Rev. Lett. \textbf{109}, 253201 (2012).
\bibitem{HAF} H. A. F\"{u}rst, N. V. Ewald, T. Secker, J. Joger, T. Feldker and R. Gerritsma, J. Phys. B: At. Mol. Opt. Phys. \textbf{51}, 195001 (2018).

\bibitem{Haze2018CDS}S. Haze, M. Sasakawa, R. Saito, R. Nakai, and T. Mukaiyama, Phys. Rev. Lett. \textbf{120}, 043401 (2018).
\bibitem{arxivKarpa} L. Karpa, arXiv:2106.06459
\bibitem{Ejtemaee20173DS} S. Ejtemaee and P. C. Haljan, Phys. Rev. Lett. \textbf{119}, 043001 (2017).

\bibitem{DeVoe2009PLD} R. G. DeVoe, Phys. Rev. Lett. \textbf{102}, 063001 (2009).
\bibitem{Rouse2017SSE}I. Rouse and S. Willitsch, Phys. Rev. Lett. \textbf{118}, 143401 (2017).







\bibitem{Leibfried} D. Leibfried, R. Blatt, C. Monroe, and D. Wineland,
Rev. Mod. Phys. \textbf{75}, 281 (2003).


\bibitem{setup} H. Hirzler, T. Feldker, H. Fürst, N. V. Ewald, E. Trimby, R. S. Lous, J. D. Arias Espinoza, M. Mazzanti, J. Joger, and R. Gerritsma, Phys. Rev. A \textbf{102}, 033109 (2020).
\bibitem{Berkeland} D. J. Berkeland, J. D. Miller, J. C. Bergquist, W. M. Itano, and D. J. Wineland, Journal of Applied Physics \textbf{83}, 5025 (1998).
\bibitem{Harter}A. Härter and J. Hecker Denschlag, Contemporary Physics \textbf{55}, 1 (2014).

\bibitem{Deb} N. Deb, L. L. Pollum, A. D. Smith, M. Keller, C. J. Rennick, B. R. Heazlewood, and T. P. Softley, Phys. Rev. A \textbf{91}, 033408 (2015).
\bibitem{KjaergaardDrewsen} N. Kjærgaard and M. Drewsen, Physics of Plasmas \textbf{8}, 1371 (2001).
\bibitem{Bandelow} S. Bandelow, G. Marx and L. Schweikhard, International Journal of Mass Spectrometry \textbf{336} (2013).

\bibitem{HasegawaBollinger} T. Hasegawa and J. J. Bollinger, Phys. Rev. A \textbf{72}, 043403 (2005).
\bibitem{rotsaddle} O. N. Kirillov and M. Levi, Am. J. Phys. \textbf{84}, 26 (2016).
\bibitem{Wester2009RMT} R. Wester, J. Phys. B: At. Mol. Opt. Phys. \textbf{42}, 154001 (2009).
\bibitem{Asvany2009NSK} O. Asvany and S. Schlemmer, International Journal of Mass Spectrometry
\textbf{279}, 2–3 (2009).
\bibitem{Notzold2020TMI} M. Nötzold, S. Z. Hassan, J. Tauch, E. Endres, R. Wester and M. Weidemüller, Appl. Sci. \textbf{10}, 5264 (2020).
\bibitem{HoltkemeierPRL} B. Höltkemeier, P. Weckesser, H. López-Carrera and M. Weidemüller, Phys. Rev. Lett.\textbf{ 116}, 233003 (2016).
\bibitem{MultiTextbook} D. Gerlich (1992) \textit{Inhomogeneous rf fields: a versatile tool for the study of processes with slow ions, Advances in Chemical Physics} (82) New York: John Wiley $\&$ Sons, Inc.
\bibitem{Schaetz} T. Huber, A. Lambrecht, J. Schmidt, L. Karpa and T. Schaetz, Nature Communications \textbf{5}, 5587 (2014).
\bibitem{Karpa} J. Schmidt, P. Weckesser, F. Thielemann, T. Schaetz and L. 
Karpa, Phys. Rev. Lett. \textbf{124}, 053402 (2020).
\bibitem{Sias} E. Perego, L. Duca and C. Sias, Appl. Sci. \textbf{10}, 2222 (2020).
\bibitem{Grimmreview} R. Grimm, M. Weidem\"{u}ller and Y. B.Ovchinnikov, Advances In Atomic, Molecular, and Optical Physics \textbf{42}, 95-170 (2000).
\bibitem{Yb+polarizability} A. Roy, S. De, Bindiya Arora and B. K. Sahoo, J. Phys. B: At. Mol. Opt. Phys. \textbf{50}, 205201 (2017).
\bibitem{TomzaPhotoa} M. Tomza, C. P. Koch, and R. Moszynski, Phys. Rev. A \textbf{91}, 042706 (2015).
\bibitem{LeBlanc2007SSO}L. J. LeBlanc and J. H. Thywissen, Phys. Rev. A \textbf{75}, 053612 (2007).
\bibitem{Lituneout} M. S. Safronova, U. I. Safronova, and Charles W. Clark, Phys. Rev. A \textbf{86}, 042505 (2012).


\bibitem{Idziaszek2007CCS} Z. Idziaszek, T. Calarco, and P. Zoller, Phys. Rev. A 76, 033409 (2007).
\bibitem{Idziaszek2011MQD} Z. Idziaszek, A. Simoni, T. Calarco, and P. S. Julienne, New J. Phys. 13, 083005 (2011).
\bibitem{Nguyen2012MMT} L. H. Nguyên, A. Kalev, M. D. Barrett, and B.-G. Englert, Phys. Rev. A \textbf{85}, 052718 (2012).
\bibitem{Joger2014QDA} J. Joger, A. Negretti, and R. Gerritsma
Phys. Rev. A \textbf{89}, 063621 (2014).

\bibitem{Gerritsma2012BJJ} R. Gerritsma, A. Negretti, H. Doerk, Z. Idziaszek, T. Calarco, and F. Schmidt-Kaler, Phys. Rev. Lett. \textbf{109}, 080402 (2012).
\bibitem{Krych2009CCT} M. Krych and Z. Idziaszek, Phys. Rev. A \textbf{80}, 022710 (2009).
\bibitem{Cook1985QTP} R. J. Cook, D. G. Shankland, and A. L. Wells, Phys. Rev. A \textbf{31}, 564 (1985).


\bibitem{Alheit1995OIP} R. Alheit, C. Hennig, R. Morgenstern, F. Vedel, and G. Werth, Applied Physics B \textbf{61} 277–283 (1995).
\bibitem{Pedregosa2010ACR} J. Pedregosa, C. Champenois, M. Houssin, and M. Knoop, Int. J. Mass Spectrom. \textbf{290}, 2–3 15 (2010).
\bibitem{DengDBS2015} K. Deng, H. Che, Y. Lan, Y. P. Ge, Z. T. Xu, W. H. Yuan, J. Zhang, and Z. H. Lu, J. Appl. Phys. \textbf{118}, 113106 (2015).

\bibitem{Pinkas2020EIT}M. Pinkas, Z. Meir, T. Sikorsky, R. Ben-Shlomi, N. Akerman, and R. Ozeri New J. Phys. \textbf{22} 013047 (2020).




\end{thebibliography}
\end{document}